%% file: pairalg.tex
\documentclass[12pt]{iopart}

\usepackage{graphicx}
\usepackage{citesort}

\usepackage{setstack} % IOP: various macros -- but \text and \substack do not work as advertised
\usepackage{amssymb}
\usepackage{amstext}  
\providecommand{\lvert}{|}
\providecommand{\rvert}{|}
\providecommand{\tfrac}[2]{{\case{#1}{#2}}}  % IOP guide page 13

\newenvironment{pseudocases}{\left\lbrace\begin{array}{r@{\quad}l}}{\end{array}\right.}

\makeatletter
\def\multline{\eqnarray\fl\def\\{\nonumber\@eqncr}}

\makeatother

\newenvironment{gathered}{\begin{array}{c}}{\end{array}}

\newenvironment{lgathered}{\begin{array}{l}}{\end{array}}
\newenvironment{pseudoaligned}{\begin{array}{r@{}l}}{\end{array}}
\newenvironment{alignedll}{\begin{array}{l@{\quad}l}}{\end{array}}

\renewcommand{\substack}[1]{{\scriptstyle\scriptsize\begin{array}{c}#1\end{array}}}

\newenvironment{ruledtabular}{\begin{indented}\item[]}{\end{indented}}

\DeclareRobustCommand{\abs}[1]{\lvert#1\rvert}
\usepackage{liealg}
\usepackage{wignert}

\usepackage{stackrel}
\DeclareRobustCommand{\subqn}[2]{\stackrel[#2]{}{#1}}

\DeclareRobustCommand{\Gaa}{G_{aa}}
\DeclareRobustCommand{\Gab}{G_{ab}}
\DeclareRobustCommand{\Gba}{G_{ba}}
\DeclareRobustCommand{\Gbb}{G_{bb}}
\DeclareRobustCommand{\GcircG}{G\circ G}
\DeclareRobustCommand{\GaacircGaa}{G_{aa}\circ G_{aa}}
\DeclareRobustCommand{\GbbcircGbb}{G_{bb}\circ G_{bb}}
\DeclareRobustCommand{\FcircF}{F\circ F}
\DeclareRobustCommand{\gqn}[2]{\stackrel[{#2}]{}{#1}}

\DeclareRobustCommand{\Gop}[2]{G^{(#2)}_{#1}}
\DeclareRobustCommand{\Fop}[1]{F^{(#1)}}

\DeclareRobustCommand{\ad}{a^\dagger}
\DeclareRobustCommand{\at}{\tilde{a}}
\DeclareRobustCommand{\bd}{b^\dagger}
\DeclareRobustCommand{\bt}{\tilde{b}}
\DeclareRobustCommand{\sd}{s^\dagger}
\DeclareRobustCommand{\st}{\tilde{s}}
\DeclareRobustCommand{\dd}{d^\dagger}
\DeclareRobustCommand{\dt}{\tilde{d}}

\DeclareRobustCommand{\Svec}{\mathrm{\mathbf{S}}}
\DeclareRobustCommand{\Jvec}{\mathrm{\mathbf{J}}}
\DeclareRobustCommand{\jhat}{\hat{\jmath}}

\DeclareRobustCommand{\ctwogrpsosp}[2][]{
\Biggl\lbrace\begin{array}{c}C_2[\grpso[#1]{#2}]\\C_2[\grpsp[#1]{#2}]\end{array}\Biggr\rbrace
}
\DeclareRobustCommand{\bbR}{\mathbb{R}}

\newcommand{\withcalgrp}[1]{{\renewcommand{\grpfont}{\mathcal}#1}}
\DeclareRobustCommand{\grpcalsu}[2][]{\withcalgrp{\grp{SU}{#1}{#2}}}

\newcommand{\eprint}[2][arXiv]{#1:\linebreak[0]#2}

\begin{document}
%bibliographystyle{apsrevm}
%bibliographystyle{iopart-num}

%***************************************************************************
% front matter
%***************************************************************************

\title{Dual algebraic structures for the two-level pairing model}

%% \author{M. A. Caprio}
%% \affiliation{Department of Physics, University of Notre Dame,
%% Notre Dame, Indiana 46556-5670, USA}
%% \author{J. H. Skrabacz}
%% \altaffiliation[Present address: ]{Department of Physics, University of California, Santa Barbara, Santa Barbara, California 93106-9530, USA}
%% \affiliation{Department of Physics, University of Notre Dame,
%% Notre Dame, Indiana 46556-5670, USA}
%% \author{F. Iachello}
%% \affiliation{Center for Theoretical Physics, Sloane Physics Laboratory, 
%% Yale University, New Haven, Connecticut 06520-8120, USA}

\author{
M A Caprio,$^1$ 
J H Skrabacz$^1$\footnote{Present address: Department of Physics, University of California, Santa Barbara, Santa Barbara, California 93106-9530, USA.},
F Iachello$^2$
}
\address{$^1$~Department of Physics, University of Notre Dame,
Notre Dame, Indiana 46556-5670, USA}
\address{$^2$~Center for Theoretical Physics, Sloane Physics Laboratory, 
Yale University, New Haven, Connecticut 06520-8120, USA}

%% \date{\today}

\begin{abstract}
Duality relations are explicitly established relating the Hamiltonians
and basis classification schemes associated with the number-conserving
unitary and number-nonconserving quasispin algebras for the two-level
system with pairing interactions.  These relations  are obtained
in a unified formulation for both bosonic and fermionic systems, with
arbitrary and, in general, unequal degeneracies for the two levels.
Illustrative calculations are carried out comparing the bosonic and
fermionic quantum phase transitions.
\end{abstract}

% 02.20.Qs -- General: Math methods: Group: General properties, structure, repn. of Lie
% 03.65.Fd -- General: Quantum mechanics: Algebraic methods

%***************************************************************************
% main text
%***************************************************************************

\section{Introduction}
\label{sec-intro}

The two-level pairing model describes a finite system which undergoes
a second-order quantum phase transition between weak-coupling and
strong-coupling dynamical symmetry limits.  This quantum phase
transition is characterized by singularities in the evolution of
various ground state properties as the pairing interaction strength
is varied: (1)~a discontinuity in the second derivative of the ground
state eigenvalue, (2)~a discontinuity in the first derivative of the
quantum order parameter, which is defined by the relative population
$\langle N_2\rangle-\langle N_1\rangle$ of the two levels and is
analogous to the magnetization parameter in the Ising model, and (3)~a
vanishing energy gap $\Delta$ between the ground state and first
excited state with the same conserved quantum numbers, and thus a
singular level density $\rho\sim\Delta^{-1}$.  Although true
singularities in these quantities only occur in the limit of infinite
particle number, ``precursors'' are found at finite $N$, which
approach the singular limit according to definite power-law
scalings~\cite{botet1983:lipkin-scaling,rowe2004:ibm-critical-scaling,dusuel2004:lipkin-scaling,dusuel2005:lipkin-scaling,dusuel2005:bcs-scaling,dusuel2005:scaling-ibm,dusuel2005:two-level-boson-cut,leyvraz2005:lipkin-scaling}.
The quantum phase transition in the two-level pairing model has long
been of interest for applications to
nuclei~\cite{hoegaasenfeldman1961:pairing-approx,broglia1968:pairing-transition,bes1970:collective-pairing-1-model}.  It has recently
served as a testbed for considering phase transitional phenomena,
including the finite-size scaling just described, excited state
quantum phase
transitions~\cite{leyvraz2005:lipkin-scaling,cejnar2006:excited-qpt,heinze2006:ibm-o6-u5-part1-level-dynamics,caprio2008:esqpt,iachello2010:qpt-carr},
thermodynamic properties~\cite{sumaryada2007:pairing-thermodynamics},
decoherence~\cite{perezfernandez2009:esqpt-decoherence}, and
quasidynamical symmetry~\cite{rowe2004:ibm-quasidynamical}, as well as
for developing the theoretical methods for treatment of these
phenomena, including continuous unitary
transformation~\cite{dusuel2005:bcs-scaling,dusuel2005:two-level-boson-cut}
and Holstein-Primakoff boson
expansion~\cite{arias2007:two-level-holstein-primakoff}.

Finite systems with pairing-type interactions, and consisting either
of bosons or fermions, occur in a broad variety of physical contexts.
Fermionic examples include superconducting grains
(electrons)~\cite{vondelft2001:supercond-grain-spectroscopy} and the
atomic nucleus (nucleons)~\cite{deshalit1963:shell}.  Bosonic examples
include the $s$-wave and $d$-wave nucleon pairs of the interacting
boson model (IBM)~\cite{iachello1987:ibm}, which themselves undergo a
bosonic pairing interaction in the description of nuclear quadrupole
collectivity, and condensates of trapped bosonic
atoms~\cite{law1998:bec-phase,uchino-2008:spinor-bec}.

The Lie algebraic properties of the two-level bosonic and fermionic
systems are closely parallel, but the differences which do arise
fundamentally affect the irreducible representations (irreps) under
which the eigenstates transform and therefore are essential to
defining the spectroscopy of the system.  Two complementary algebraic
formulations are relevant to the description of finite pairing
systems~\cite{kerman1961:pairing-collective,helmers1961:shell-sp,judd1968:group-atomic,moshinsky1970:noninvariance-groups,wybourne1974:groups,rowe2010:rowanwood}:
a \textit{unitary} algebra is spanned by the bilinear products of a
creation and anniliation
operator~\cite{racah1949:complex-spectra-part4-f-shell,racah1965-group-spectro},
and a \textit{quasispin}
algebra~\cite{kerman1961:pairing-collective,kerman1961:pairing-approx,ui1968:su11-quasispin-shell}
is defined in terms of creation and annihilation operators for
time-reversed pairs of particles.  These structures are intertwined by
duality relations, in particular, relating irreps of the quasispin
algebra with those of an orthogonal (in the case of bosons) or
symplectic (in the case of fermions) subalgebra of the full unitary
algebra.  Such relations have often been
used~\cite{lawson1965:quasispin-me,macfarlane1966:shell,arima1966:quasispin-shell,arima1979:ibm-o6,pan1998:so6u5,volya2001:pairing-quasispin,rowe2010:rowanwood}
to effect simplifications of the calculations for two-level and
multi-level systems.

In this article, the duality relations between the unitary and
quasispin algebraic structures for the two-level system are
systematically established.  In particular, attempts to compare
results across two-level systems with different level degeneracies or
between the bosonic and fermionic cases (see
Ref.~\cite{caprio2008:esqpt}) raise the question as to which
differences in spectroscopic results are
\textit{superficial}, \textit{i.e.}, originating from an imperfect
choice of correspondence between the Hamiltonian parameters for the
two cases, and which are due to more \textit{fundamental} or
irreconciliable distinctions.  Therefore, a main intent of the present
work is to resolve the relationships between the disparate forms of
the Hamiltonian which arise in the definitions of the dynamical
symmetries and in numerical studies of the transition between them.
These Hamiltonians include: (1)~the Casimir form defined in terms of the unitary
algebra, (2)~the pairing form used in studies of the fermionic system,
which is essentially defined in terms of quasispin operators, and
(3)~the ``multipole'' form traditionally considered for physical
reasons in bosonic studies.  The relations are established in a fully
general fashion which uniformly accomodates arbitary level
degeneracies ($n_1$ and $n_2$), for both the bosonic and fermionic
cases.  

Previous work on two-level systems has concentrated either on the
so-called $s$-$b$ boson models, with level degeneracies $n_1=1$
and $n_2\geq1$ (\textit{i.e.}, for which one of the levels is a
singlet), or on fermionic models of equal degeneracies
($n_1=n_2$). The observations outlined here are intended to provide a
foundation for more detailed future work, allowing for the most
general choice of level degeneracies.  The results are provided as a
basis for algebraic studies of the quantum phase transitions, excited
state spectroscopic structure, and classical
geometry~\cite{gilmore1974:lie-groups,feng1981:ibm-phase} of two-level
and multi-level pairing models.  Although the discussion is presented
for two-level systems, for the sake of clarity, many of the results
carry over to multi-level systems essentially without modification.

After a brief summary of the dual
algebraic structures for the many-body problem in general
(Sec.~\ref{sec-bose-fermi}), the unitary algebraic structure is
presented in detail, including categorization of the subalgebra
structure, classification of the irreducible representations,
construction of the generators, and identification of the Casimir
operators, all in a unified form for bosonic and fermionic cases
(Sec.~\ref{sec-unitary}).  The simpler quasispin structure is also reviewed
(Sec.~\ref{sec-quasi}).  Duality relations are then established
between the unitary (or Casimir) and quasispin (or pairing) formulations of
the Hamiltonian (Sec.~\ref{sec-hamiltonian}).  These are explicitly
related to the spectral properties of the two-level system through
numerical calculations across the quantum phase transition,
illustrating basic distinctions between the bosonic and fermionic
cases, when calculated for bosonic and fermionic systems with with similar level degeneracies and/or similar particle
number (Sec.~\ref{sec-trans}).

\section{Bosonic and fermionic algebras}
\label{sec-bose-fermi}

The fundamental Lie algebra describing transformations of a many-boson or
many-fermion system is spanned by the bilinear products of creation
and/or annihilation operators  $\ad_{m'}
\ad_{m}$, $\ad_{m'} a_{m}$, and $a_{m'} a_{m}$ (\textit{e.g.},
Refs.~\cite{wybourne1974:groups,rowe2010:rowanwood}).  For bosons, the
resulting algebra is $\grpsp{2n,\bbR}$, and for fermions it is
$\grpso{2n}$, where $m$ and $m'=1$, $\ldots$, $n$ range over the single
particle states of the system.  Two important sets of subalgebras
arise: number-\textit{conserving} subalgebras and
number-\textit{nonconserving} (quasispin) subalgebras.

The restriction to number-conserving operators, spanned by the
elementary one-body operators $\ad_{m'} a_{m}$, constitutes a
$\grpu{n}$ algebra.  The $\grpu{n}$ algebra contains a subalgebra
$\grpso{n}$ for the bosonic case or $\grpsp{n}$ for the fermionic
case.  If each single-particle creation operator $\ad_{m}$ is
associated with a time-reversed partner $\ad_{\bar{m}}$, these
$\grpso{n}$ or $\grpsp{n}$ subalgebras are defined by the property
that they leave invariant the ``scalar'' pair state $\sum_m
\ad_{m}\ad_{\bar{m}}\tket{0}$~\cite{rowe2010:rowanwood}.  This special
property underlies the duality relations with the quasispin pair
algebra considered in the present work.  More specifically, we
consider rotationally-invariant problems, for which the
single-particle states may be identified as the $2j+1$ substates of
single-particle levels of various angular momenta $j$ (\textit{i.e.},
$j$-shells, in the nomenclature of nuclear physics, which we adopt for
either bosonic or fermionic levels).  Then the creation operators are
of the form $\ad_{m}\rightarrow\ad_{km}$ and
$\ad_{\bar{m}}\rightarrow(-)^{j_k-m}\ad_{k,-m}$ for the $k$th level.  For such
rotationally-invariant systems, the $\grpso{n}$ or $\grpsp{n}$
subalgebras in turn contain the physical $\grpso{3}\sim\grpsu{2}$
angular momentum algebra.  Although we follow the convention
of denoting the angular momentum algebra $\grpso{3}$ in the bosonic
case and $\grpsu{2}$ in the fermionic case, there is no material distinction
between the algebras.  In general, there may also be other,
intervening subalgebras in the chain.

Alternatively, the scalar pair creation operator $S_+=\tfrac12\sum
\ad_{m}\ad_{\bar{m}}$, the scalar pair annihilation operator $S_-=\tfrac12\sum
a_{\bar{m}}a_{m}$, and the number-conserving operator $S_0=\tfrac14\sum(
\ad_{m}a_{m}+\theta a_m \ad_{m})$ close under commutation, where
$\theta=+$ for bosonic systems or $\theta=-$ for fermionic systems.
These operators define a
number-nonconserving pair quasispin algebra, either $\grpcalsu{1,1}$ for
bosons or  $\grpcalsu{2}$ for fermions.
%%\footnote{
The caligraphic
notation for the quasispin
algebras is adopted~\cite{rowe2010:rowanwood} to avoid ambiguity between
the $\grpcalsu{2}$ quasispin algebra
and the $\grpsu{2}$ angular momentum
algebra.
%%}

In summary, for a bosonic system, the subalgebras under consideration are
\begin{equation}
\label{eqn-chain-full-bose}
\grpsp{2n,\bbR}\supset
\left\lbrace
\begin{array}{l}
\grpu{n}\supset\grpso{n}\supset\cdots\supset\grpso{3}\\
\grpcalsu{1,1}\supset\grpu{1},
\end{array}
\right.
\end{equation}
and, for a fermionic system, they are
\begin{equation}
\label{eqn-chain-full-fermi}
\grpso{2n}\supset
\left\lbrace
\begin{array}{l}
\grpu{n}\supset\grpsp{n}\supset\cdots\supset\grpsu{2}\\
\grpcalsu{2}\supset\grpu{1}.
\end{array}
\right.
\end{equation}
The subalgebras of $\grpu{n}$ are useful in the classification of
states not only for the pairing Hamiltonian (defined in Sec.~\ref{sec-model}) but
also for a much richer range of
Hamiltonians~\cite{iachello2006:liealg}.  

A close relation between unitary chain and quasispin subalgebras
arises since the quasispin and orthogonal or symplectic algebras may
be embedded as mutually commuting ``dual'' algebras within the larger
$\grpsp{2n,\bbR}$ or $\grpso{2n}$ algebra.  The algebraic foundations
are discussed in detail in
Refs.~\cite{kerman1961:pairing-collective,helmers1961:shell-sp,judd1968:group-atomic,moshinsky1970:noninvariance-groups,wybourne1974:groups,rowe2010:rowanwood}.
Here we simply note that duality denotes the situation in which the
states within a space may be classified simultaneously in terms of two
mutually commuting groups (or algebras) $G_1$ and $G_2$, such that the
irrep labels of the two groups are in one-to-one correspondence.  For
the present problem, the embedding and associated labels are given,
for the bosonic case, by
\begin{equation}
\grpsp{2n,\bbR}
\supset 
\bigl[\subqn{\grpso{n}}{\lbrack v \rbrack}\supset\cdots\supset\subqn{\grpso{3}}{J}\bigr]
\otimes
\bigl[\subqn{\grpcalsu{1,1}}{S}\supset\subqn{\grpu{1}}{N}\bigr]
\end{equation}
and, for the fermionic case, by
\begin{equation}
\grpso{2n}
\supset 
\bigl[\subqn{\grpsp{n}}{\lbrace v\rbrace}\supset\cdots\supset\subqn{\grpsu{2}}{J}\bigr]
\otimes
\bigl[\subqn{\grpcalsu{2}}{S}\supset\subqn{\grpu{1}}{N}\bigr].
\end{equation}
The seniority label $v$ (Sec.~\ref{sec-unitary-branch}) and quasispin
label $S$ (Sec.~\ref{sec-quasi}) are in one-to-one correspondence,
\textit{i.e.}, specifying the value of one uniquely determines the
value of the other, and \textit{vice versa}.

The duality relations hold equally well regardless of whether the
$n$-dimensional single-particle space is construed to consist of a
single $j$-shell ($n=2j+1$, odd for bosons or even for fermions), two
$j$-shells ($n=n_1+n_2$), or, indeed, multiple $j$-shells.  The
one-level case has been considered in detail (\textit{e.g.},
Ref.~\cite{wybourne1974:groups}).  However, we find that the detailed
construction of operators for two-level systems within the context of
this duality, as needed for spectroscopic studies of these systems,
requires elaboration.  Although, for simplicity, we consider only the
case of two levels, the results may readily be generalized to
additional levels.

\section{Unitary algebra}
\label{sec-unitary}

\subsection{Subalgebra chains}
\label{sec-unitary-sub}

Consider the $\grpu{n}$ subalgebra
chains for the two-level system, consisting of either bosonic or fermionic
levels, of possibly unequal degeneracies.  If the levels are
$j$-shells of angular momenta $j_1$ and $j_2$, the level degeneracies
are $n_1=2j_1+1$ and $n_2=2j_2+1$, and the total degeneracy of the
system is $n=n_1+n_2$.  For the bosonic case, we have
\begin{multline}
%% for AMS: \renewcommand{\arraystretch}{\subalgarraystretch}
\label{eqn-chainunbose}
\gqn{\grpu{n_1+n_2}}{[N]}
\supset
\left\lbrace
\begin{array}{c}
\gqn{\grpso{n_1+n_2}}{[v]}\\
\gqn{\grpu[1]{n_1}}{[N_1]}\otimes\gqn{\grpu[2]{n_2}}{[N_2]}
\end{array}
\right\rbrace
\nonumber\\
\supset
\gqn{\grpso[1]{n_1}}{[v_1]}\otimes\gqn{\grpso[2]{n_2}}{[v_2]}
\supset
\gqn{\grpso[1]{3}}{J_1}\otimes\gqn{\grpso[2]{3}}{J_2}
\supset
\gqn{\grpso[12]{3}}{J} 
\end{multline}
and, for the fermionic case, we have
\begin{multline}
\label{eqn-chainunfermi}
\gqn{\grpu{n_1+n_2}}{\lbrace N\rbrace}
\supset
\left\lbrace
\begin{array}{c}
\gqn{\grpsp{n_1+n_2}}{\lbrace v\rbrace}\\
\gqn{\grpu[1]{n_1}}{\lbrace N_1 \rbrace}\otimes\gqn{\grpu[2]{n_2}}{\lbrace N_2\rbrace}
\end{array}
\right\rbrace
\nonumber\\
\supset
\gqn{\grpsp[1]{n_1}}{\lbrace v_1\rbrace}\otimes\gqn{\grpsp[2]{n_2}}{\lbrace v_2\rbrace}
\supset
\gqn{\grpsu[1]{2}}{J_1}\otimes\gqn{\grpsu[2]{2}}{J_2}
\supset
\gqn{\grpsu[12]{2}}{J} ,
\end{multline}
where $n_1=2j_1+1$ and $n_2=2j_2+1$.  The irreducible representation
labels, indicated beneath the symbol for each algebra, are defined in
Sec.~\ref{sec-unitary-branch}, and the algebras themselves are constructed explicitly in
Sec.~\ref{sec-unitary-gen}.  Throughout the following discussion,
bosonic and fermionic cases will be considered in parallel.

The subalgebras summarized in~(\ref{eqn-chainunbose})
and~(\ref{eqn-chainunfermi}) are generically present, regardless of
the level degeneracies $n_1$ and $n_2$, for $n_1$ and $n_2\geq2$.
However, several clarifying comments are in order:

(1) The important special case of a singlet bosonic level ($j_k=0$) leads to
$n_k=1$, and the corresponding orthogonal algebra $\grpso[k]{n_k}$ is
undefined.  The label $v_k$ may still be defined, in a limited sense,
through the quasispin, as noted in Sec.~\ref{sec-quasi}.  Two-level
bosonic problems in which $j_1=0$ are termed $s$-$b$ boson models.
These include the Schwinger boson realization ($j_1=j_2=0$) of the
Lipkin model~\cite{lipkin1965:lipkin-model-part1-exact-pert}.  The
subalgebra chains and labeling schemes for the $s$-$b$ models were
considered in Ref.~\cite{caprio2008:esqpt}.

(2) For a fermionic level with $j_k=\tfrac12$, and therefore $n_k=2$,
the symplectic algebra $\grpsp[k]{n_k}$ in~(\ref{eqn-chainunfermi}) is
identical to the $\grpsu[k]{2}$ angular momentum algebra.

(3) Additional subalgebras of $\grpu{n_1+n_2}$ may also arise, parallel to chains indicated
above and still
containing the angular momentum algebra,
\textit{e.g.}, for
the interacting boson model ($n_1=1$ and $n_2=5$), there is a
physically relevant chain
$\grpu{6}\supset\grpsu{3}\supset\grpso[12]{3}$~\cite{iachello1987:ibm}.
However, since these chains are not directly relevant to the pairing
problem and cannot be treated in a uniform fashion for arbitrary $n_1$
and $n_2$, they are not considered further here.

(4) Further subalgebras may also intervene between $\grpso{n}$ and
$\grpso{3}$, or between $\grpsp{n}$ and $\grpsu{2}$, the classic
example being the appearance of the exceptional algebra $\mathrm{G}_2$ in the chain
$\grpso{7}\supset\mathrm{G}_2\supset\grpso{3}$~\cite{racah1949:complex-spectra-part4-f-shell}.

(5) Whenever two realizations of the same algebra commute with each
other, the sum generators $G_i=G^{(1)}_i+G^{(2)}_i$ also form a new
realization of the algebra, as in ordinary angular momentum
addition.  For instance, addition of
the angular momentum generators for the two levels
[$L=L^{(1)}+L^{(2)}$] gives the generators of the sum angular momentum
algebra $\grpso[12]{3}$ (bosonic) or
$\grpsu[12]{2}$ (fermionic)
in~(\ref{eqn-chainunbose}) and~(\ref{eqn-chainunfermi}).  However,
more generally, if the two levels have equal degeneracies
($n_1=n_2\equiv n_{12}$), such a combination of generators may also be
made higher in the subalgebra chains~(\ref{eqn-chainunbose})
and~(\ref{eqn-chainunfermi}), yielding
$\grpu[1]{n_{12}}\otimes\grpu[2]{n_{12}}\supset\grpu[12]{n_{12}}\supset\grpso[12]{3}$
and
$\grpso[1]{n_{12}}\otimes\grpso[2]{n_{12}}\supset\grpso[12]{n_{12}}\supset\grpso[12]{3}$
for the bosonic case, or similarly
$\grpu[1]{n_{12}}\otimes\grpu[2]{n_{12}}\supset\grpu[12]{n_{12}}\supset\grpsu[12]{2}$
and
$\grpsp[1]{n_{12}}\otimes\grpsp[2]{n_{12}}\supset\grpsp[12]{n_{12}}\supset\grpsu[12]{2}$
for the fermionic case.

\subsection{Branching}
\label{sec-unitary-branch}

The branching rules for the irreps arising in the bosonic or fermionic
realizations of the algebras in~(\ref{eqn-chainunbose})
and~(\ref{eqn-chainunfermi}) provide the classification of states for
the two-level pairing model.  Some, but not all, of these branchings
can be expressed in closed form.

For the {\it bosonic} realization of $\grpu{n}$,
the symmetric irreps $[N]\equiv[N0\ldots0]$ (with $n$ labels) are
obtained, where $N$ is the occupation number.  For $\grpso{n}$, the
irreps are $[v]\equiv[v0\ldots0]$ (with $\lfloor n/2 \rfloor$ labels).

The $\grpu{n}\rightarrow\grpso{n}$ branching is of the type considered by
Hammermesh~\cite{hammermesh:group} and is given by
\begin{equation}
\label{eqn-branchuso}
v=(N \bmod 2), \ldots, N-2, N.
\end{equation}
This rule applies both to the branching
$\grpu{n_1+n_2}\rightarrow\grpso{n_1+n_2}$ and to the branching
associated with each of the two levels in
$\grpu[1]{n_1}\otimes\grpu[2]{n_2}\rightarrow\grpso[1]{n_1}\otimes\grpso[2]{n_2}$.
Note that $n$ is odd for a single bosonic level and is even for the
two-level system.  [The
$\grpu{n}\rightarrow\grpu[1]{n_1}\otimes\grpu[2]{n_2}$ branching
rule follows trivially from additivity of the number operators,
$N=N_1+N_2$.]

For the branching
$\grpso{n_1+n_2}\rightarrow\grpso[1]{n_1}\otimes\grpso[2]{n_2}$, the
allowed $v_1$ and $v_2$ are obtained by considering all partitions of
$v$ as
\begin{equation}
\label{eqn-branchsoprod}
v=v_1+v_2+2n_v \quad (n_v=0,1,\ldots,\lfloor v/2\rfloor).
\end{equation}
This rule may be verified by dimension counting arguments, that is,
$\dim [v]=\sum_{v_1v_2}\dim[v_1]\dim[v_2]$, using the Weyl dimension
formula~\cite{hammermesh:group}.  Notice that for the bosonic system (in contrast to the
fermionic case below) the branching rule for
$\grpso{n_1+n_2}\rightarrow\grpso[1]{n_1}\otimes\grpso[2]{n_2}$ is
independent of the level degeneracies $n_1$ and
$n_2$, and the total occupation number $N$
influences the allowed $\grpso[1]{n_1}$ and $\grpso[2]{n_2}$ irreps
only through the constraint~(\ref{eqn-branchuso}) on $v$.  

The allowed partitions $v\rightarrow(v_1v_2)$ are listed, for low $v$,
in Table~\ref{tab-branch-partition}.  As a concrete example, for $N=2$, the allowed
$\grpso{n_1+n_2}$ irreps have $v=0$ and $2$, with branchings to
$\grpso[1]{n_1}\otimes\grpso[2]{n_2}$ given by the corresponding rows of
Table~\ref{tab-branch-partition}.  As a specific example of the
equivalence of dimensions, consider the case of the 
two-level bosonic system with $j_1=j_2=1$, thus described by
$\grpso{6}\supset\grpso[1]{3}\otimes\grpso[2]{3}$.  The $v=2$ irrep of
$\grpso{6}$ has
dimension $20$, while the corresponding $\grpso{3}\otimes\grpso{3}$
irreps likewise have total dimension $\dim (2,0) + \dim (1,1)+\dim
(0,2)+\dim (0,0)=(5)(1)+(3)(3)+(1)(5)+(1)(1)=20$.
%--------------------------------
\begin{table}
\caption{
Partitions $v\rightarrow(v_1,v_2)$ of the form permitted by
the $\grpso{n_1+n_2}\rightarrow\grpso[1]{n_1}\otimes\grpso[2]{n_2}$
branching rule~(\ref{eqn-branchsoprod}).  The partitions permitted by
the $\grpsp{n_1+n_2}\rightarrow\grpsp[1]{n_1}\otimes\grpsp[2]{n_2}$
branching rule~(\ref{eqn-branchspprod}) are a subset of these.
}
\label{tab-branch-partition}
\input{pairalg_tab01.tex}
\end{table}
%--------------------------------

The branchings of the form $\grpso{n}\rightarrow\grpso{3}$, needed for
$\grpso[1]{n_1}\otimes\grpso[2]{n_2}\rightarrow\grpso[1]{3}\otimes\grpso[2]{3}$,
are more complicated and, in general, involve missing labels.
However, such branchings occur widely in physical applications, and
general methods exist for the solution based on weights or character
theory~\cite{judd1963:operator-techniques,wybourne1974:groups}.  An
explicit multiplicity formula is obtained in
Ref.~\cite{gheorghe2004:so-branching}, applicable to the symmetric
irreps arising in the present bosonic case~(\ref{eqn-chainunbose}).
Finally, the reduction
$\grpso[1]{3}\otimes\grpso[2]{3}\rightarrow\grpso[12]{3}$ is governed
by the usual triangle inequality for angular momentum addition.

The branching rules for the {\it fermionic} case are nearly identical,
with a few modifications.  For $\grpu{n}$ we obtain the {\it
anti}\/symmetric irreps $\lbrace N \rbrace \equiv
[1,\ldots,1,0,\ldots,0]$, that is, with $N$ unit entries (out of $n$
labels total).  Similarly, for $\grpsp{n}$ we have $\lbrace v \rbrace \equiv
[1,\ldots,1,0,\ldots,0]$, with $v$ unit entries (out of $n/2$ labels total).

The branching rule for $\grpu{n}\rightarrow\grpsp{n}$~\cite{hammermesh:group} requires the
modification of~(\ref{eqn-branchuso}) to
\begin{equation}
\label{eqn-branchusp}
%% \tag{\ref{eqn-branchuso}$'$}
v=(N' \bmod 2), \ldots, N'-2, N',
\end{equation}
where $N'\equiv\min(N,n-N)$.  Notice, therefore, that
$v\leq\tfrac12n$.  This rule applies to
$\grpu{n_1+n_2}\rightarrow\grpsp{n_1+n_2}$ and to
$\grpu[1]{n_1}\otimes\grpu[2]{n_2}\rightarrow\grpsp[1]{n_1}\otimes\grpsp[2]{n_2}$.

The values of $v_1$ and $v_2$ arising in the branching
$\grpsp{n_1+n_2}\rightarrow\grpsp[1]{n_1}\otimes\grpsp[2]{n_2}$ are again
given by the partitioning condition~(\ref{eqn-branchsoprod}), but now
subject to an additional constraint, so
\begin{equation}
\label{eqn-branchspprod}
%% \tag{\ref{eqn-branchsoprod}$'$}
%% \Biggl\lbrace
\begin{array}{l}
%%\begin{gathered}
v=v_1+v_2+2n_v \quad (n_v=0,1,\ldots,\lfloor v/2\rfloor)\\
\abs{(v_1-v_2)-\tfrac12(n_1-n_2)}\leq \tfrac12(n_1+n_2)-v
,
%%\end{gathered}
\end{array}
\end{equation}
as can
again be verified by dimensional counting.  
The branching rules~(\ref{eqn-branchusp}) and~(\ref{eqn-branchspprod})
together automatically enforce $v_1\leq\tfrac12n_1$ and
$v_2\leq\tfrac12n_2$.  For two levels of {\it equal} degeneracy
($n_1=n_2\equiv n_{12}$), the constraint simplifies to
\begin{equation}
\label{eqn-branchspprodequal}
\abs{v_1-v_2}\leq n_{12}-v
\end{equation}
and only serves to exclude $(v_1,v_2)$ values when $v>\tfrac12n_{12}$.

For illustration, branchings for the low-dimensional case
$\grpsp{4}\rightarrow\grpsp{2}\otimes\grpsp{2}$ (two $j=\tfrac12$
levels) are given in Table~\ref{tab-branch-sp4}.  The chain
$\grpsp{4}\supset\grpsp{2}\otimes\grpsp{2}$ is isomorphic to the
canonical chain
$\grpso{5}\supset\grpso{4}$ of orthogonal algebras, and the
branchings given in Table~\ref{tab-branch-sp4} therefore also follow from
the $\grpso{n}$ canonical branching
rule~\cite{gelfand1950:repn-ortho,hecht1965:so5-wigner,kemmer1968:so5-irreps-1}.
The $\grpso{5}$ Cartan labels are given by 
$[\lambda_1,\lambda_2]=[\tfrac12(\lambda_1'+\lambda_2'),\tfrac12(\lambda_1'-\lambda_2')]$,
where $[\lambda_1',\lambda_2']$ are the $\grpsp{4}$ Cartan labels, and
the $\grpso{4}$ Cartan labels are given by 
$[\lambda_1,\lambda_2]=[\tfrac12(v_1+v_2),\tfrac12(v_1-v_2)]$ (see
Ref.~\cite{caprio2010:racah} for a summary of notation for this chain). 
Branchings for
the higher-dimensional case
$\grpsp{20}\supset\grpsp{10}\otimes\grpsp{10}$ are given in
Table~\ref{tab-branch-sp20}.  
%--------------------------------
\begin{table}
\caption{
Branching $\grpsp{4}\rightarrow\grpsp[1]{2}\otimes\grpsp[2]{2}$
for irreps of
the two-level fermionic system consisting of two $j=1/2$ levels.
Allowed irreps are a subset of those listed in
Table~\ref{tab-branch-partition}.  The Cartan labels of these irreps
with respect to the 
isomorphic chain $\grpso{5}\supset\grpso{4}$ are also indicated.}
\label{tab-branch-sp4}
\input{pairalg_tab02.tex}
\end{table}
%--------------------------------
%--------------------------------
\begin{table}
\caption{
Branching $\grpsp{20}\rightarrow\grpsp[1]{10}\otimes\grpsp[2]{10}$
for irreps of
the two-level fermionic system consisting of two $j=9/2$ levels, with $N=10$.
The fermionic branching rule~(\ref{eqn-branchspprod}) restricts the
allowed irreps, relative to those listed in
Table~\ref{tab-branch-partition}, for $v>\tfrac12n_{12}=5$.  }
\label{tab-branch-sp20}
\input{pairalg_tab03.tex}
\end{table}
%--------------------------------

\subsection{Generators}
\label{sec-unitary-gen}

Of the generators for the algebras in chains~(\ref{eqn-chainunbose})
and~(\ref{eqn-chainunfermi}), those involving a single level
are well known~\cite{racah1965-group-spectro}.  Here we must 
construct the generators for the two-level system.
Since the subalgebra chains terminate in the two-level angular
momentum algebra,
$\grpso[12]{3}\sim\grpsu[12]{2}$, it is most natural to express the
generators as spherical tensors with respect to this angular momentum
algebra.

First, let us briefly review the results for the single $j$-shell,
with creation operators $\ad_m$ and annihilation operators $a_m$
($m=-j$, $-j+1$, $\ldots$, $j$).  In the case of a single {\it
bosonic} level, with degeneracy $n=2j+1$ ($j$ integer), the subalgebra
chain for $\grpu{n}$ is $\grpu{n} \supset \grpso{n} \supset
\cdots\supset\grpso{3}$ [see~(\ref{eqn-chain-full-bose})].  The
generators of $\grpu{n}$, in spherical tensor form, are the bilinears
\begin{equation}
G^{(g)}_\gamma=(\ad\times\at)^{(g)}_\gamma \qquad(g=0,1,\ldots,2j),
\end{equation}
where $\gamma=-g$, $-g+1$, $\ldots$, $+g$.
The product of two spherical tensor operators is defined by
$(A^a\times B^b)^c_\gamma=\sum_{\alpha\beta}\tcg a\alpha b\beta
c\gamma A^a_\alpha B^b_\beta$.  We  follow the time reversal phase
convention
$\tilde{A}^a_\alpha=(-)^{a-\alpha}A^a_{-\alpha}$~\cite{varshalovich1988:am}.
Thus, \textit{e.g.}, $\at_{m}=(-)^{j-m}a_{-m}$, where the time
reversal phase factor is required for the annihilation operator to
transform as a spherical tensor under rotation.\footnote{The
convention $\tilde{A}^{(a)}_\alpha=(-)^{a+\alpha}A^{(a)}_{-\alpha}$
also arises in the literature.  The relative sign between these
conventions implies straightforward modifications
$\tilde{A}^{(a)}\rightarrow (-)^{2a}\tilde{A}^{(a)}$ to signs throughout the
following results.}

The commutators of the generators are most conveniently expressed in the spherical
tensor  coupled
form (see appendix)
\begin{equation}
\label{eqn-Gcomm}
\bigl[ G^{(e)}, G^{(f)} \bigr]^{(g)} 
=
 (-)^g [1-(-)^{e+f+g}] \hat{e} \hat{f}
\smallsixj{e}{f}{g}{j}{j}{j}G^{(g)},
\end{equation}
where we adopt the shorthand $\jhat=(2j+1)^{1/2}$.
The coefficient on the right hand side
of~(\ref{eqn-Gcomm}) vanishes unless $e+f+g$ is odd.  Consequently,
the generators $G^{(g)}_\gamma$ with $g$ odd ($g=1$, $3$, $\ldots$,
$2j-1$) close under commutation, forming the basis for the subalgebra
$\grpso{n}$.  Finally, the generators $G^{(1)}_\gamma$, which span  
the 
$\grpso{3}$ algebra,  are
proportional to the physical angular momentum generators
\begin{math}
L^{(1)}_\lambda =
[\tfrac13j(j+1)(2j+1)]^{1/2}
(\ad\times\at)^{(1)}_\lambda
\end{math}
for a single bosonic $j$-shell.  

For a single {\it fermionic} level, with degeneracy $n=2j+1$ ($j$ half-integer),
we have instead the chain $\grpu{n} \supset \grpsp{n} \supset \cdots \supset
\grpsu{2}$ [see~(\ref{eqn-chain-full-fermi})].
The generators $G^{(g)}_\gamma$ of $\grpu{n}$ again obey the
commutation relations~(\ref{eqn-Gcomm}), and the generators with $g$
odd ($g=1$, $3$, $\ldots$, $2j$) now span the $\grpsp{n}$ algebra.  The
$G^{(1)}_\gamma$ are proportional to the physical angular momentum
operators, now given by
\begin{math}
L^{(1)}_\lambda =-
[\tfrac13j(j+1)(2j+1)]^{1/2}
(\ad\times\at)^{(1)}_\lambda,
\end{math}
closing as an $\grpsu{2}$ algebra.

Proceeding now to the algebras involving both levels of the two-level
system, let us reduce the complexity of the subscripts, relative to
the generic multi-level notation $\ad_{km}$, by denoting the
creation operators for the two levels by $\ad_\alpha(\equiv \ad_{1,\alpha})$ and $\bd_\beta(\equiv \ad_{2,\beta})$,
respectively, with angular momenta $j_a\equiv j_1$ and $j_b\equiv j_2$, where $\alpha=-j_a$,
$-j_a+1$, $\ldots$, $+j_a$ and $\beta=-j_b$, $-j_b+1$,
$\ldots$, $+j_b$.  The level degeneracies appearing in the
algebra labels are $n_1=2j_a+1$ and $n_2=2j_b+1$.
The algebra
$\grpu{n_1+n_2}$ is spanned by
\begin{equation}
\label{eqn-Gkkdefn}
\begin{alignedll}
\Gaa^{(g)}= (\ad\times\at)^{(g)} & (g=0,1,\ldots,2j_a)
\\ %%\quad&
\Gab^{(g)}= (\ad\times\bt)^{(g)} & (g=\abs{j_a-j_b},\ldots,j_a+j_b)
\\
\Gba^{(g)}= (\bd\times\at)^{(g)} & (g=\abs{j_a-j_b},\ldots,j_a+j_b)
\\ %%\quad&
\Gbb^{(g)}= (\bd\times\bt)^{(g)} & (g=0,1,\ldots,2j_b).
\end{alignedll}
\end{equation}
The commutation relations for these generators are listed in
Table~\ref{tab-comm-un}.  They may all be obtained from the general
bilinear commutation relation~(\ref{eqn-comm-bilinear}).  Notice the
nearly identical commutation relations for the bosonic and fermionic
realizations of the $\grpu{n_1+n_2}$ algebra, with sign differences
indicated by the presence of the symbol $\theta$ in
Table~\ref{tab-comm-un} (recall that $\theta=+$ for the bosonic case
and $\theta=-$ for the fermionic case).  Commutators not listed in
Table~\ref{tab-comm-un}, \textit{e.g.},
$[\Gab^{(e)},\Gba^{(f)}]^{(g)}$, can be obtained from those given, by
the coupled commutator symmetry relation~(\ref{eqn-comm-symm}).  The
subalgebra $\grpu[1]{n_1}\otimes\grpu[2]{n_2}$ is obtained by simply
omitting the ``mixed'' generators $\Gab^{(g)}$ and $\Gba^{(g)}$.
%--------------------------------
\begin{table}
\caption{Commutation relations for the generators of the $\grpu{n_1+n_2}$ algebra of the 
two-level \textit{bosonic} system ($\theta=+$) or \textit{fermionic}
system ($\theta=-$), in coupled form.}
\label{tab-comm-un}
\input{pairalg_tab04.tex}
\end{table}
%--------------------------------
%--------------------------------
\begin{table}
\caption{Commutation relations for the generators of the $\grpso{n_1+n_2}$ algebra of the 
two-level \textit{bosonic} system ($\theta=+$) or the
$\grpsp{n_1+n_2}$  algebra of the \textit{fermionic}
system ($\theta=-$), in coupled form.}
\label{tab-comm-sosp}
\input{pairalg_tab05.tex}
\end{table}
%--------------------------------

The $\grpso{n_1+n_2}$ subalgebra, for the bosonic case, or
$\grpsp{n_1+n_2}$ subalgebra, for the fermionic case, is then obtained by
restricting~(\ref{eqn-Gkkdefn}) to the following generators:
$\Gaa^{(g)}$ with $g$ odd ($g=1$, $3$, $\ldots$, $2j_a-1$ or $2j_a$),
$\Gbb^{(g)}$ with $g$ odd ($g=1$, $3$, $\ldots$, $2j_b-1$ or $2j_b$),
and certain linear combinations of the form
\begin{math}
F^{(g)}=\eta_g\Gab^{(g)}+\xi_g\Gba^{(g)},
\end{math}
the coefficients of which are determined by the requirement of closure.
Specifically, using the results of Table~\ref{tab-comm-un}, it is
found that closure is obtained if $\xi_g/\eta_g=\sigma_0(-)^g$ for
all $g$.  That
is, the relative sign between the terms must alternate, between
generators $F^{(g)}$ with even and odd
tensor rank $g$, but an overall sign parameter $\sigma_0$ may be
chosen as either $\pm1$.  Thus, we obtain
\begin{equation}
\label{eqn-Feta}
F^{(g)}=\eta_g[(\ad\times\bt)^{(g)} +\sigma_0(-)^g(\bd\times\at)^{(g)}].
\end{equation}
Note therefore that there are actually {\it two} distinct
$\grpso{n_1+n_2}$ subalgebras which may be included
in~(\ref{eqn-chainunbose}), or two $\grpsp{n_1+n_2}$ subalgebras in
chain~(\ref{eqn-chainunfermi}), distinguished by the relative sign
$\sigma_0$ in the generators $F^{(g)}$.

An overall arbitrary phase $\eta_g$ remains in the definition of $F^{(g)}$.
If we choose
$F^{(g)}$ to be a ``self-adjoint'' tensor, in the sense that
\begin{equation}
F^{(g)\,\dagger}=\tilde{F}^{(g)},
\end{equation}
\textit{i.e.},
\begin{math}
F^{(g)\,\dagger}_\gamma=(-)^{g-\gamma}F^{(g)}_{-\gamma},
\end{math}
the commutation relations among the generators take on a simple form
and, moreover, involve only real coefficients.
Let $\sigma_0=(-)^s$, with $s=0$ or $1$.  Then a self-adjoint tensor $F^{(g)}$
is obtained for the choice
\begin{equation}
\label{eqn-Fphase}
\eta_g=
\begin{pseudocases}
 1 & \text{$a+b+s$ even}\\
 i & \text{$a+b+s$ odd}.
\end{pseudocases}
%% \cases{1 & $a+b+s$ even\\i & $a+b+s$ odd.\\}
\end{equation}
The commutation relations for the $\grpso{n_1+n_2}$ or
$\grpsp{n_1+n_2}$ generators are listed in
Table~\ref{tab-comm-sosp}.

The phase choice~(\ref{eqn-Fphase}) for the two-level generator $F^{(g)}$ also offers
consistency with the $s$-$b$ boson models, where $F^{(j_b)}$ plays an
important role as a physical transition operator.  For instance, in
the IBM ($j_s=0$ and $j_d=2$), the choice $\sigma_0=+$ (\textit{i.e.},
$s=0$) gives
$\grpso{6}$ generator
\begin{math}
F^{(2)}=(\sd\times\dt)^{(2)}+(\dd\times\st)^{(2)},
\end{math}
which is the leading-order electric quadrupole operator~\cite{arima1979:ibm-o6}.  The choice
$\sigma_0=-$ (\textit{i.e.}, $s=1$) instead yields the generator 
\begin{math}
F^{(2)}=i[(\sd\times\dt)^{(2)}-(\dd\times\st)^{(2)}]
\end{math}
of a distinct $\grpso{6}$ subalgebra, denoted by
$\overline{\grpso{6}}$~\cite{vanisacker1985:ibm-so6}, which has been
shown to be relevant to the decomposition of nuclear excitations into
intrinsic and collective parts~\cite{leviatan1987:ibm-intrinsic}.

Finally,  construction of the remaining subalgebras in~(\ref{eqn-chainunbose})
and~(\ref{eqn-chainunfermi}) follows by application of the
same principles.  The
$\grpso[1]{n_1}\otimes\grpso[2]{n_2}$ or
$\grpsp[1]{n_1}\otimes\grpsp[2]{n_2}$ algebra is obtained by
restriction to $\Gaa^{(g)}$ and $\Gbb^{(g)}$ with $g$ odd, and $\grpso[1]{3}\otimes\grpso[2]{3}$ or
$\grpsu[1]{2}\otimes\grpsu[2]{2}$ is obtained by further restriction to
$g=1$.  The combined angular momentum algebra, $\grpso[12]{3}$ or
$\grpsu[12]{2}$, then has generators $L^{(1)}_\lambda$, where
\begin{equation}
L^{(1)} =\theta
[\tfrac13j_a(j_a+1)(2j_a+1)]^{1/2}
\Gaa^{(1)}
+\theta
[\tfrac13j_b(j_b+1)(2j_b+1)]^{1/2}
\Gbb^{(1)}
.
\end{equation}

\subsection{Casimir operators}
\label{sec-unitary-casimir}

To exploit the symmetry properties of the two-level pairing model with
respect to the subalgebras of $\grpu{n_1+n_2}$, it will be necessary
(Sec.~\ref{sec-unitary-dynsymm}) to express the Hamiltonian in terms
of the quadratic Casimir operators of the algebras
in~(\ref{eqn-chainunbose}) and~(\ref{eqn-chainunfermi}).
Identification of the Casimir operator proceeds in two stages.  First,
a quadratic operator which commutes with the generators must be
identified.  This only defines the Casimir operator to within a
normalization (and phase) factor.  It is then desirable to choose the
normalization such that the eigenvalues of the Casimir operator match
the conventional eigenvalue
formulas~\cite{nwachuku1977:so-sp-casimir,iachello2006:liealg}, given
in terms of the Cartan highest-weight labels for the irrep in
Table~\ref{tab-casimir}.  For the symmetric irreps of $\grpso{n}$ or
antisymmetric irreps of $\grpsp{n}$ arising in the two-level pairing
problem, the eigenvalues can be expressed in terms of the single
unified formula
\begin{equation}
\label{eqn-eigen-c2sosp}
\biggl\langle\ctwogrpsosp{n}\biggr\rangle=2v(\theta v + n -2\theta).
\end{equation}
Thus, as the second stage of defining the Casimir operator, the
normalization  is evaluated by explicitly considering the action of
the operator on the one-body states, for which the irrep labels are known.
%--------------------------------
\begin{table}
% unindent caption and unindent table while still using indented
% environment for rest of formatting (font size, etc.)
\setlength{\mathindent}{0pt} 
\caption{Casimir operator eigenvalue formulas relevant to the
two-level pairing model.  Expressions are given first in terms of the
generic Cartan irrep labels $[\lambda_1\lambda_2\ldots\lambda_k]$ and
then specialized to the specific irreps arising for the two-level
pairing system.}
\label{tab-casimir}
\input{pairalg_tab06.tex}
\end{table}
%--------------------------------

For the \textit{single-level} algebra, $\grpso{n}$ or $\grpsp{n}$, the
operator $\GcircG$, defined by\footnote{The generators of $\grpso{n}$
or $\grpsp{n}$ together transform as an $\grpso{n}$ or $\grpsp{n}$
tensor.  Therefore, the
circle in the notation $\GcircG$ is meant to represent a scalar
product with respect to $\grpso{n}$ or $\grpsp{n}$, following
Ref.~\cite{caprio2007:geomsuper2}.  In~(\ref{eqn-FcircFdefn}) the notation is
generalized to represent an $\grpso[1]{n_1}\otimes\grpso[2]{n_2}$ or
$\grpsp[1]{n_1}\otimes\grpsp[2]{n_2}$ scalar. }
\begin{equation}
\label{eqn-GcircGdefn}
\GcircG=-\sum_{\text{$g$ odd}} \hat{g} [G^{(g)}\times
G^{(g)}]^{(0)}_0,
\end{equation}
commutes with all the generators, \textit{i.e.}, $G^{(g)}$ with $g$
odd~\cite{racah1949:complex-spectra-part4-f-shell,judd1963:operator-techniques,uchino-2008:spinor-bec}.
The result follows from the general theory of Casimir operators for an
algebra, and it may be verified by explicitly evaluating the
commutator $[G^{(g)},\GcircG]^{(g)}$, using
commutator~(\ref{eqn-Gcomm}) and product
rule~(\ref{eqn-coupled-product-rule}).  In terms of the conventional
spherical tensor scalar product, defined by $A^{(c)}\cdot B^{(c)}=
(-)^c \hat{c} (A^{(c)}\times B^{(c)})^{(0)}_0$, this operator is
\begin{math}
\GcircG=\sum_{\text{$g$ odd}} G^{(g)}\cdot
G^{(g)}.
\end{math}
The eigenvalue of $\GcircG$ acting on the one-body state $\tket{a}$
may easily be evaluated by Wick's theorem in coupled
form~\cite{chen1993:wick-coupled}, using the commutator results
described in the appendix.  Comparison with the eigenvalue
formula~(\ref{eqn-eigen-c2sosp}) gives normalization
\begin{equation}
\label{eqn-C2sosp}
\ctwogrpsosp{n}=4\GcircG,
\end{equation}
covering both the bosonic and fermionic cases.

Proceeding to the \textit{two-level} problem, for the Casimir operator of
$\grpso{n_1+n_2}$ or $\grpsp{n_1+n_2}$, we start from the Casimir
operators $4\GaacircGaa$ and $4\GbbcircGbb$ of each single-level
subalgebra, as defined in~(\ref{eqn-GcircGdefn}).  Although each of
these operators commutes with each of the generators of
$\grpso[1]{n_1}\otimes\grpso[2]{n_2}$ or
$\grpsp[1]{n_1}\otimes\grpsp[2]{n_2}$, they do not commute with the
two-level generators $F^{(g)}$.  We therefore introduce the
operator
\begin{equation}
\label{eqn-FcircFdefn}
\FcircF=\sum_g\hat{g}[F^{(g)}\times F^{(g)}]^{(0)}_0,
\end{equation}
or, equivalently,
\begin{math}
\FcircF=\sum_g(-)^g F^{(g)}\cdot F^{(g)},
\end{math}
with $g=\abs{j_a-j_b},\ldots,j_a+j_b$.  This quantity is invariant with respect to
$\grpso[1]{n_1}\otimes\grpso[2]{n_2}$ or
$\grpsp[1]{n_1}\otimes\grpsp[2]{n_2}$.  Moreover, the combination
\begin{equation}
\label{eqn-C2sospab}
\ctwogrpsosp{n_1+n_2}
=2\theta\FcircF+4\GaacircGaa +4\GbbcircGbb
\end{equation}
commutes with all the generators of $\grpso{n_1+n_2}$ or
$\grpsp{n_1+n_2}$, as seen by application of the commutators
in Table~\ref{tab-comm-sosp} and the product
rule~(\ref{eqn-coupled-product-rule}).  That this combination of
operators also has the
correct normalization to match the eigenvalue formula for
$C_2[\grpso{n}]$ ($n$ even) or $C_2[\grpsp{n}]$
(Table~\ref{tab-casimir}) may be verified by explicitly calculating the
one-body expectation value $\tme{a}{C_2}{a}$ or
$\tme{b}{C_2}{b}$.

Returning to the example of the IBM $\grpso{6}\supset\grpso{5}$ chain, the Casimir
operator~(\ref{eqn-C2sosp}) becomes 
\begin{equation}
\label{eqn-c2so5}
C_2[\grpso{5}]=4(\dd\times\dt)^{(1)}\cdot(\dd\times\dt)^{(1)} 
+4(\dd\times\dt)^{(3)}\cdot(\dd\times\dt)^{(3)},
\end{equation}
for the single
level consisting of the quadrupole boson $d^{(2)}$.
Then, for the $\grpso{6}$ algebra of the two-level $s$-$d$ system, 
\begin{equation}
\label{eqn-c2so6}
C_2[\grpso{6}]=2[(\sd\times\dt)^{(2)}+(\dd\times\st)^{(2)}]
\cdot[(\sd\times\dt)^{(2)}+(\dd\times\st)^{(2)}]+C_2[\grpso{5}],
\end{equation}
consistent with the usual result~\cite{iachello1987:ibm}.

Similar results apply to the quadratic Casimir
operators of the unitary algebras in~(\ref{eqn-chainunbose})
and~(\ref{eqn-chainunfermi}).  For the \textit{single-level} algebra, the linear
invariant of $\grpu{n}$ is simply the occupation number
operator $N=\sum_m \ad_{m}a_{m}$, or $N=\theta\jhat G^{(0)}_0$.  The
quadratic invariant is given by
\begin{equation}
\label{eqn-C2un}
C_2[\grpu{n}]=\sum_g \hat{g} (-)^g [G^{(g)}\times
G^{(g)}]^{(0)}_0,
\end{equation}
or 
\begin{math}
C_2[\grpu{n}]=\sum_g G^{(g)}\cdot
G^{(g)}.
\end{math}
However, for the bosonic realization of $\grpu{n}$, only symmetric
irreps arise, and, for the fermionic realization of $\grpu{n}$, only antisymmetric
irreps arise, with eigenvalues as given in 
Table~\ref{tab-casimir}.  Therefore, in either situation, it is found
that the
quadratic invariant is simply a function of the linear invariant and can
be expressed as 
\begin{equation}
C_2[\grpu{n}]=N(\theta N+n-\theta).
\end{equation}
Likewise, the Casimir operator of the two-level system's algebra
$\grpu{n_1+n_2}$ may be expressed as
\begin{equation}
\label{eqn-C2unab}
C_2[\grpu{n_1+n_2}]=2\theta N_1N_2+ n_2N_1+n_1N_2 
+C_2[\grpu[1]{n_1}]+C_2[\grpu[2]{n_2}].
\end{equation}
This result is obtained by comparison of the eigenvalues for
$C_2[\grpu{n_1+n_2}]$ with those for $C_2[\grpu{n_1}]$ and
$C_2[\grpu{n_2}]$, together with the additivity of the number
operators ($N=N_1+N_2$).

\section{Quasispin algebra}
\label{sec-quasi}

First, we note that a set of three operators $S_{+}$, $S_{-}$, and $S_{z}$
obeying the commutation relations
\begin{equation}
\label{eqn-Scomm}
[S_{0},S_{+}]=+S_{+}
\quad [S_{0},S_{-}]=-S_{-}
\quad
[S_{+},S_{-}]=-2\theta S_{0},
\end{equation}
and obeying the  unitarity conditions $S_{+}^\dagger=S_{-}$ and
$S_0^\dagger=S_0$, span a unitary realization either of the algebra
$\grpsu{1,1}$, for $\theta=+$, or of the algebra $\grpsu{2}$,
for $\theta=-$.  The $\grpsu{1,1}$ or $\grpsu{2}$ invariant operator
is given by
\begin{equation}
\label{eqn-Ssqr}
\begin{pseudoaligned}
\Svec^2&=S_{0}^2-\tfrac12\theta(S_{+}S_{-}+S_{-}S_{+})\\
&=S_{0}(S_{0}-1) -\theta S_{+}S_{-}.
\end{pseudoaligned}
\end{equation}
For an irrep of $\grpsu{1,1}$, this operator takes on eigenvalues
$S(S-1)$, and the possible eigenvalues of $S_0$ are given by $M=S$,
$S+1$, $S+2$, $\ldots$.  For the ``true'' group
representations of $\grpsu{1,1}$, $S$ must be integer or half-integer,
but description of the bosonic system in the quasi-spin formalism as
considered below requires the projective representations with
$S=\tfrac14$, $\tfrac34$, $\tfrac54$, $\ldots$, for reasons described
in Ref.~\cite{ui1968:su11-quasispin-shell}.  As usual, for
$\grpsu{2}$, $\Svec^2$ takes on eigenvalues $S(S+1)$, with $S=0$,
$\tfrac12$, $1$, $\ldots$, and the eigenvalues of $S_0$ are given by
$M=-S$, $\ldots$, $+S-1$, $+S$.

Now, consider a system consisting of one or more $j$-shells of angular momentum
$j_k$ ($k=1$, $2$, $\ldots$).  Regardless of whether the operators
$\ad_{km}$ for the $k$th level are
bosonic or fermionic, a quasispin algebra is defined following the
prescription of Sec.~\ref{sec-bose-fermi}.  The scalar pair creation operator, scalar pair annihilation
operator, and an operator simply related to the number operator for
this level form a
closed set under commutation.  Specifically, let 
\begin{equation}
\label{eqn-Sdefn}
\begin{lgathered}
S_{k+}=\tfrac12\sum_m\ad_{km}\ad_{k\bar{m}} 
\\%%\quad
S_{k-}=\tfrac12\sum_m a_{k\bar{m}}a_{km} 
\\%%\quad
S_{k0}=\tfrac14\sum_m(\ad_{km}a_{km} +\theta a_{km}\ad_{km}).
\end{lgathered}
\end{equation}
These define either an $\grpsu{1,1}$ quasispin algebra for
bosons~\cite{ui1968:su11-quasispin-shell}, which we denote by $\grpcalsu[k]{1,1}$, or an $\grpsu{2}$ quasispin
algebra for
fermions~\cite{kerman1961:pairing-collective,helmers1961:shell-sp},
which we denote by $\grpcalsu[k]{2}$. In
terms of spherical tensor coupled products, 
these operators~(\ref{eqn-Sdefn}) may be represented as
\begin{equation}
\label{eqn-Stensor}
\begin{lgathered}
S_{k+}=\tfrac12\jhat_k(\ad_k\times\ad_k)^{(0)}_0
\\%%\quad
S_{k-}=\tfrac12\theta\jhat_k(\at_k\times\at_k)^{(0)}_0
\\%%\quad
S_{k0}=\tfrac14\jhat_k\theta[(\ad_k\times\at_k)^{(0)}_0+(\at_k\times\ad_k)^{(0)}_0].
\end{lgathered}
\end{equation}
The operator $S_{k0}$ is related to the number operator
$N_k=\sum_m\ad_{km}a_{km}$, which may be expressed in spherical tensor form as
$N_k=\jhat_k(\ad_k\times\at_k)^{(0)}_0$,
by
\begin{equation}
\label{eqn-Sznumber}
S_{k0}=\tfrac12(N_k+\theta\Omega_k),
\end{equation}
where the constant $\Omega_k=\tfrac12(2j_k+1)$ is the pair degeneracy of
the level $k$.  The quasispin $S_k$, moreover, is related to the
seniority quantum number $v_k$ by the duality relation for a single
$j$-shell.

It is therefore possible to interconvert between quasispin quantum
numbers $S_k$ and $M_k$, for each level, and occupation-seniority
quantum numbers $N_k$ and $v_k$, according to
\begin{equation}
\label{eqn-SMreln}
\begin{lgathered}
S_k=\tfrac12(\Omega_k+\theta v_k)\\
M_k=\tfrac12(N_k+\theta\Omega_k).
\end{lgathered}
\end{equation}
Since the lowest weight state for a given quasispin (\textit{i.e.},
with $M_k=\theta S_k$) is destroyed by the
pair annihilation operator, and since this state contains $N_k=v_k$
particles, $v_k$ may be interpreted as the number of unpaired
particles, either bosons or fermions.  The seniority $v_k$ takes on
values $v_k=0$, $1$, $2$, $\ldots$, subject to the constraint $v_k\leq
N_k$ for bosons
or $v_k\leq
\min(N_k,2\Omega_k-N_k)$ for fermions, by the $M$ contents of
the irreducible representations noted above.\footnote{As noted in
Sec.~\ref{sec-unitary-sub}, the case of a $j=0$ bosonic level is
anomalous.  There is no orthogonal algebra dual to the quasispin
algebra, and thus no seniority quantum number, but the label $v_k$ may
still be defined from the quasispin via~(\ref{eqn-SMreln}).  The
squared quasispin operator is identically $\Svec^2=-\tfrac3{16}$ for
such a level, as described in
Refs.~\cite{lipkin1966:lie-pedestrians,ui1968:su11-quasispin-shell}.
Thus $S_k=\tfrac14$ or $\tfrac34$, and hence $v_k=0$ or $1$.  Since
$M-S$ must be integral, it follows that $v_k=0$ for $N_k$ even and
$v_k=1$ for $N_k$ odd,
\textit{i.e.}, $v_k= N_k \bmod 2$.  The natural interpretation
of this value is that
particles in a $j=0$ level are automatically paired to zero angular
momentum, except for the one unpartnered particle when the occupation
is odd.}
Note, therefore, in the fermionic case, that
$v_k\leq\Omega_k$, with the maximum value occuring for half filling
($N_k=\Omega_k$).

A quasispin algebra for the two-level system~--- which we
denote by
$\grpcalsu[12]{1,1}$ for the bosonic case or $\grpcalsu[12]{2}$ for the
fermionic case~--- is spanned by the sum generators
\begin{equation}
\label{eqn-S12}
S_+=S_{1+}+\sigma S_{2+} \quad S_-=S_{1-}+\sigma S_{2-} \quad S_0=S_{10}+ S_{20}.
\end{equation}
A quasispin algebra is obtained with either choice of sign $\sigma=\pm$
in the ladder operators.
This algebra defines a total quasispin quantum number $S$ which is
dual to the two-level algebra seniority quantum number $v$, by a
relation of the same form as~(\ref{eqn-SMreln}), namely,
\begin{equation}
\label{eqn-SMreln-sum}
\begin{lgathered}
S=\tfrac12(\Omega+\theta v)\\
M=\tfrac12(N+\theta\Omega),
\end{lgathered}
\end{equation}
where $N$ and $\Omega$ are defined above as the sums of the single-level values.
The allowed values for the total quasispin $S$ are given for
$\grpcalsu[12]{1,1}$ by $S\geq S_1+S_2$ (\textit{i.e.}, $S_1\otimes
S_2\rightarrow S_1+S_2$, $S_1+S_2+1$, $\ldots$) and for
$\grpcalsu[12]{2}$ by the familiar triangle inequality $\abs{S_1-S_2}\leq
S\leq S_1+S_2$ (\textit{i.e.}, $S_1\otimes
S_2\rightarrow \abs{S_1-S_2}$, $\ldots$, $S_1+S_2-1$, $S_1+S_2$).  

When reexpressed in terms of the seniority labels
$v_1$, $v_2$, and $v$, the $\grpcalsu{1,1}$ coupling rule is equivalent
to the $\grpso{n_1+n_2}\rightarrow\grpso[1]{n_1}\otimes\grpso[2]{n_2}$
branching rule~(\ref{eqn-branchsoprod}), and the $\grpcalsu{2}$ coupling
rule is equivalent to the
$\grpsp{n_1+n_2}\rightarrow\grpsp[1]{n_1}\otimes\grpsp[2]{n_2}$
branching rule~(\ref{eqn-branchspprod}).  Similarly, the $\grpcalsu{1,1}$
$M$ content rule is equivalent to the
$\grpu{n}\rightarrow\grpso{n}$ branching rule~(\ref{eqn-branchuso}),
and the $\grpcalsu{2}$ $M$ content rule is equivalent to the
$\grpu{n}\rightarrow\grpsp{n}$ branching rule~(\ref{eqn-branchusp}).

\section{Hamiltonian relations}
\label{sec-hamiltonian}

\subsection{Dynamical symmetries}
\label{sec-unitary-dynsymm}

Before considering the pairing Hamiltonian in particular, let us
consider the Hamiltonian defined by the Casimir operators of the
unitary subalgebra chains.  A dynamical symmetry~\cite{iachello2006:liealg}
arises when the Hamiltonian is constructed in terms of the Casimir
operators of a \textit{single} chain of subalgebras.  The eigenstates thus
reduce the subalgebra chain, \textit{i.e.}, constitute
irreps of the subalgebras.  More generally,
especially when considering phase transitions, it is useful to
construct the Hamiltonian from terms consisting of Casimir operators
from multiple, parallel chains, here the upper and lower chains
of~(\ref{eqn-chainunbose}) or~(\ref{eqn-chainunfermi}), as
\begin{multline}
\label{eqn-H-casimir}
H=aN+b_1N_1+b_2N_2+b\ctwogrpsosp{n_1+n_2}
\nonumber\\
+c_1\ctwogrpsosp[1]{n_1}+c_2\ctwogrpsosp[2]{n_2}
+d_1\Jvec_1^2+d_2\Jvec_2^2+e\Jvec^2,
\end{multline}
where higher-order invariants may also be included

The upper chain in~(\ref{eqn-chainunbose})
or~(\ref{eqn-chainunfermi}) defines an $\grpso{n_1+n_2}$ or
$\grpsp{n_1+n_2}$ dynamical symmetry, and the lower chain
defines a
$\grpu[1]{n_1}\otimes\grpu[2]{n_2}$ dynamical symmetry.  
The $\grpu[1]{n_1}\otimes\grpu[2]{n_2}$ dynamical symmetry is obtained
for the Hamiltonian~(\ref{eqn-H-casimir}) with $b=0$, \textit{i.e.},
\begin{multline}
\label{eqn-H-UxU}
H=aN+b_1N_1+b_2N_2
\nonumber\\
+c_1\ctwogrpsosp[1]{n_1}+c_2\ctwogrpsosp[2]{n_2}
+d_1\Jvec_1^2+d_2\Jvec_2^2+e\Jvec^2.
\end{multline}
 The eigenstates
$\tket{NN_1N_2v_1v_2\cdots J_1J_2J}$ have definite occupation numbers for each of the
levels and have energy eigenvalues
\begin{multline}
\label{eqn-E-UxU}
E=aN+b_1N_1+b_2N_2+2c_1 v_1(\theta v_1+n_1-2\theta)+2c_2 v_2(\theta
v_2+n_2-2\theta)
\nonumber\\
+d_1 J_1(J_1+1)+d_2 J_2(J_2+1)+e J(J+1).
\end{multline}

\begin{figure}
\begin{center}
\includegraphics*[width=0.8\hsize]{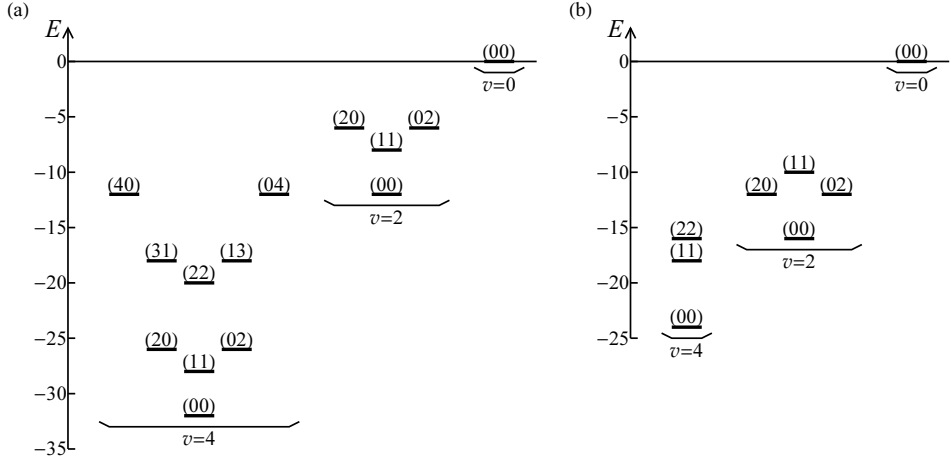}
\end{center}
\caption{Energy diagrams for the bosonic $\grpso{n_1+n_2}$ and
fermionic $\grpsp{n_1+n_2}$ dynamical symmetries of the two-level
pairing model. The degeneracy within irreducible representations of
the two-level algebra (brackets labeled by $v$) is split according to
the single-level algebra irreducible representation labels
$(v_1v_2)$. (a)~Energy levels for
the bosonic $\grpso{6}\supset\grpso{3}\otimes\grpso{3}$ dynamical symmetry
Hamiltonian ($j_1=j_2=1$).  (b)~Energy levels for
the fermionic $\grpsp{8}\supset\grpsp{4}\otimes\grpsp{4}$ dynamical symmetry
Hamiltonian ($j_1=j_2=\tfrac32$).  The Hamiltonian
in each case is chosen as $H=-\theta\FcircF$, \textit{i.e.},
$b=-\tfrac12$ and
$c_1=c_2=+\tfrac12$ in~(\ref{eqn-H-sosp}).  The total
occupation in both cases is $N=4$, which for the fermionic example
gives half filling.
}
\label{fig-schemes}
\end{figure}

The $\grpso{n_1+n_2}$ or $\grpsp{n_1+n_2}$ dynamical symmetry is
obtained for the Hamiltonian~(\ref{eqn-H-casimir}) with $b_1=b_2=0$,
\textit{i.e.},
\begin{multline}
\label{eqn-H-sosp}
H=aN+b\ctwogrpsosp{n_1+n_2}
\nonumber\\
+c_1\ctwogrpsosp[1]{n_1}+c_2\ctwogrpsosp[2]{n_2}
+d_1\Jvec_1^2+d_2\Jvec_2^2+e\Jvec^2,
\end{multline}
with eigenstates
$\tket{Nvv_1v_2\cdots J_1J_2J}$ and energy eigenvalues
\begin{multline}
\label{eqn-E-sosp}
E=aN+2b v(\theta v+n_1+n_2-2\theta) +2c_1 v_1(\theta v_1+n_1-2\theta)+2c_2 v_2(\theta
v_2+n_2-2\theta)
\nonumber\\
+d_1 J_1(J_1+1)+d_2 J_2(J_2+1)+e J(J+1).
\end{multline}
The energy spectrum for the
dynamical symmetry follows from the branching rules of
Sec.~\ref{sec-unitary-branch}.  Example level energy diagrams are
shown for a bosonic system ($n_1=n_2=3$) in Fig.~\ref{fig-schemes}(a)
and for a fermionic system of similar degeneracies ($n_1=n_2=4$) in
Fig.~\ref{fig-schemes}(b).
%--------------------------------
%--------------------------------

\subsection{Pairing Hamiltonian}
\label{sec-model}

The pairing Hamiltonian for a
generic multi-level system, consisting of levels of angular momentum $j_1$,
$j_2$, $\ldots$, is given by
\begin{equation}
\label{eqn-Hsum}
H=\sum_{km}\varepsilon_k \ad_{km} a_{km}
+\tfrac14\sum_{\substack{k'k\\m'm}}G_{k'k}(-)^{j_{k'}-m'}\ad_{k'm'}\ad_{k',-m'}(-)^{j_{k}-m}a_{k,-m}a_{km},
\end{equation}
where the summation indices $k$ and $k'$ run over the single-particle
levels ($k=1$, $2$, $\ldots$), and $m$ and $m'$ run over their
substates ($m=-j_k$, $-j_k+1$, $\ldots$, $+j_k$).  The creation
operators $\ad_{km}$ and annihilation operators $a_{km}$ obey either
canonical commutation relations (bosons) or anticommutation relations
(fermions).  The first term
represents the one-body energy contribution for each level, and
the products in the second term are creation or annihilation operators for
pairs involving time-reversed partner substates.  
Note that $G_{k'k}=G_{kk'}^*$ by Hermiticity of $H$, and that
we take all coefficients to be real.
That this Hamiltononian is specifically constructed from angular
momentum zero pair operators is seen by rewriting it in terms of
angular-momentum coupled products, as
\begin{equation}
\label{eqn-Hscalar}
H=\sum_{k}\varepsilon_k N_k
+\tfrac14\theta\sum_{k'k}G_{k'k}\jhat_{k'}\jhat_k(\ad_{k'}\times\ad_{k'})^{(0)}_0(\at_{k}\times\at_{k})^{(0)}_0.
\end{equation}

The Hamiltonian~(\ref{eqn-Hsum}) is integrable and may be solved using
a generalized Gaudin algebra~\cite{ortiz2005:gaudin-models} or, under
certain conditions, Bethe
ansatz~\cite{balantekin2007:pairing-bethe-ansatz,pan2009:nuclear-bethe-ansatz}
methods.  Two limiting cases deserve special mention, since they are
characterized by dynamical symmetries
(Sec.~\ref{sec-unitary-dynsymm}), as described below, and are solvable
by more elementary methods. (1)~Trivially, when all $G_{k'k}=0$, the
problem reduces to that of a system of noninteracting particles
(weak-coupling limit), with eigenstates characterized by occupations
numbers $N_k$.  (2)~When all $\varepsilon_k=0$, physically
corresponding to the situation in which the
level energy difference is negligible relative to the pairing strength
(strong-coupling limit), and if all $G_{k'k}$ are equal to within
possible phase factors (uniform pairing), the problem is immediately
solvable by the use of quasispin (Sec.~\ref{sec-H-quasi}).

\subsection{Quasispin Hamiltonian}
\label{sec-H-quasi}

The generic multi-level pairing Hamiltonian~(\ref{eqn-Hsum}) can be
expressed entirely in terms of the quasispin generators, by comparison
with~(\ref{eqn-Stensor}), as
\begin{equation}
\label{eqn-H-quasi-gen}
H=\sum_{k}\varepsilon_k (2S_{k0}-\theta\Omega_k)
+\sum_{k'k}G_{k'k}S_{k'+}S_{k-}.
\end{equation}
This Hamiltonian therefore conserves the quasispin $S_k$ or,
equivalently, the seniority $v_k$, associated with each level.
The Hamiltonian is also number conserving, so the total $z$-projection
$S_0=\sum_kS_{k0}=\tfrac12(N+\theta\Omega)$ is conserved, where
$N=\sum_kN_k$ and $\Omega=\sum_k\Omega_k$.  However, the individual
$S_{k0}$ are not in general conserved, unless the levels completely
decouple, with $G_{k'k}=0$ for all $k'\neq k$.

Note that numerical diagonalization is straightforward in the weak-coupling
basis, consisting of states
$\tket{N_1 v_1 N_2 v_2\cdots}$ of good seniority and occupation for
each level.  The action of the Hamiltonian~(\ref{eqn-H-quasi-gen}) on
these states follows from the known action of the quasispin ladder
operators, $S_\pm\tket{SM}=[-\theta(S\pm\theta M)(S\mp \theta M-\theta)]^{1/2}\tket{S(M\pm1)}$,
once the quantum numbers are translated
via~(\ref{eqn-SMreln}).  Specifically,
\begin{multline}
\label{eqn-Saction}
\begin{lgathered}
S_{k+}\tket{\cdots N_k v_k\cdots}
=\tfrac12 [\theta(N_k-v_k+2)(N_k+v_k+2\theta\Omega_k)]^{1/2}
\tket{\cdots (N_k+2) v_k\cdots}\\
S_{k-}\tket{\cdots N_k v_k\cdots}
=\tfrac12 [\theta(N_k-v_k)(N_k+v_k+2\theta\Omega_k-2)]^{1/2}
\tket{\cdots (N_k-2) v_k\cdots}.
\end{lgathered}
\end{multline}
Consequently, the matrix elements for the diagonal pairing terms
are
\begin{multline}
\label{eqn-SkSk-action}
\tbra{\cdots N_k v_k\cdots}
S_{k+}S_{k-}
\tket{\cdots N_k v_k\cdots}
=
\tfrac14\theta[N_k(N_k+2\theta\Omega_k-2)-v_k(v_k+2\theta\Omega_k-2)]
\end{multline}
and for the off-diagonal terms are
\begin{multline}
\label{eqn-SkpSk-action}
\tbra{\cdots (N_{k'}+2)v_{k'}\cdots(N_k-2) v_k \cdots}
S_{k'+}S_{k-}
\tket{\cdots N_{k'}v_{k'} \cdots N_k v_k \cdots}
\nonumber\\=
\tfrac14[
(N_{k'}-v_{k'}+2)
(N_{k'}+v_{k'}+2\theta\Omega_{k'})
\nonumber\\
\qquad\qquad\times
(N_{k}-v_{k})
(N_{k}+v_{k}+2\theta\Omega_{k}-2)
]^{1/2},
\end{multline}
as noted for the fermionic case in, \textit{e.g.},
Refs.~\cite{broglia1968:pairing-transition,volya2001:pairing-quasispin}.

Returning to the two-level problem, the pairing Hamiltonian in
quasispin notation is
\begin{multline}
\label{eqn-H-quasi}
H=
\varepsilon_1(2S_{10}-\theta\Omega_1)+G_{11}S_{1+}S_{1-}
+\varepsilon_2(2S_{20}-\theta\Omega_2)+G_{22}S_{2+}S_{2-}
\nonumber\\
+G_{12}(S_{1+}S_{2-}+S_{2+}S_{1-}).
\end{multline}
The two-level pairing Hamiltonian has two dynamical
symmetries~\cite{chen1990:tlpm-qpt}
defined with respect to the quasispin algebras, corresponding to either the
upper or lower subalgebra chains in
\newcommand{\subalgarraystretch}{1.3}
\begin{equation}
\label{eqn-chainsu11}
\renewcommand{\arraystretch}{\subalgarraystretch}
\gqn{\grpcalsu[1]{1,1}}{v_1}\otimes\gqn{\grpcalsu[2]{1,1}}{v_2}
\supset
\left\lbrace
\begin{array}{c}
\gqn{\grpcalsu[12]{1,1}}{v}\\
\gqn{\grpu[1]{1}}{N_1}\otimes\gqn{\grpu[2]{1}}{N_2}
\end{array}
\right\rbrace \supset \gqn{\grpu[12]{1}}{N}
\end{equation}
for the bosonic case or
\begin{equation}
\label{eqn-chainsu2}
\gqn{\grpcalsu[1]{2}}{v_1}\otimes\gqn{\grpcalsu[2]{2}}{v_2}
\supset
\left\lbrace
\begin{array}{c}
\gqn{\grpcalsu[12]{2}}{v}\\
\gqn{\grpu[1]{1}}{N_1}\otimes\gqn{\grpu[2]{1}}{N_2}
\end{array}
\right\rbrace \supset \gqn{\grpu[12]{1}}{N}
\end{equation}
for the fermionic case, with conserved quantum numbers as indicated.
Here the algebra $\grpu[k]{1}$ is the trivial Abelian algebra spanned
by $S_{k0}$, and $\grpu[12]{1}$ is spanned by the their sum $S_0$, as defined
in~(\ref{eqn-S12}).  The occupation-seniority labels ($v$ and $N$) are
indicated, rather than the quasispin labls ($S$ and $M$), for a closer
connection to the physical problem and easier comparison with the dual
algebra's dynamical symmetries.

The dynamical symmetry Hamiltonian for the upper subalgebra chain is
the strong coupling limit (\textit{i.e.},
$\varepsilon_1=\varepsilon_2=0$) of the two-level pairing Hamiltonian,
with uniform pairing strength, as defined in Sec.~\ref{sec-model}.
Specifically, let $G_{11}=\sigma G_{12}=\sigma G_{21}=G_{22}\equiv G$,
for either sign $\sigma=\pm$.  Then the Hamiltonian is given by
\begin{equation}
\label{eqn-H-quasi-equal}
H=G S_+ S_-, 
\end{equation}
where $S_\pm$ are the sum-quasispin ladder operators
of~(\ref{eqn-S12}), defined in terms of the same sign $\sigma$.
Since  $S_+S_-=\theta[S_{0}(S_{0}-1) -\Svec^2]$, 
by~(\ref{eqn-Ssqr}), the strong-coupling Hamiltonian conserves the
total quasispin $S$ (or seniority $v$), as well as the projection
quantum number $M$ (or occupation $N$), and has eigenvalues
\begin{equation}
\label{eqn-SS-eigen}
\langle S_{+}S_{-}\rangle=
\tfrac14\theta[N(N+2\theta\Omega-2)-v(v+2\theta\Omega-2)],
\end{equation}
as expressed in terms of the occupation-seniority labels.  The
eigenstates are identical to those of the $\grpso{n_1+n_2}$ or
$\grpsp{n_1+n_2}$ dynamical symmetry
Hamiltonian~(\ref{eqn-H-sosp}).\footnote{More precisely, the
quasispin Hamiltonian~(\ref{eqn-H-quasi-equal}) has a higher
degeneracy than the $\grpso{n_1+n_2}$ or
$\grpsp{n_1+n_2}$ Hamiltonian~(\ref{eqn-H-sosp}), but the
eigenstates can be chosen from within each degenerate subspace to match
those of~(\ref{eqn-H-sosp}), \textit{i.e.}, of good $J_1$, $J_2$, and $J$.} 
The specific relationship between the Hamiltonians is determined
below in Sec.~\ref{sec-dual}.

The dynamical symmetry Hamiltonian for the lower subalgebra chain is
the weak-coupling limit of the pairing Hamiltonian ($G_{k'k}=0$), as
defined in Sec.~\ref{sec-model}. The dynamical symmetry eigenstates
are simply the level occupation eigenstates of good $N_1$ and $N_2$,
as for the $\grpu[1]{n_1}\otimes\grpu[2]{n_2}$ dynamical symmetry
Hamiltonian~(\ref{eqn-H-UxU}), \textit{i.e.}, the weak-coupling basis
states considered above.

The full
two-level pairing Hamiltonian~(\ref{eqn-H-quasi}) can be expressed entirely in terms of
the invariant operators of algebras appearing in the upper and lower
chains. Specifically,
\begin{multline}
\label{eqn-H-quasi-invariant}
H=
\varepsilon_1(2S_{10}-\theta\Omega_1)
+(G_{11}-\sigma G_{12})\theta[S_{10}(S_{10}-1)-\Svec_1^2]
\nonumber\\
+\varepsilon_2(2S_{20}-\theta\Omega_2)
+(G_{22}-\sigma G_{12})\theta[S_{20}(S_{20}-1)-\Svec_2^2]
\nonumber\\
+G_{12}\sigma\theta[S_{0}(S_{0}-1)-\Svec^2].
\end{multline}

\subsection{Duality relations for the Hamiltonian}
\label{sec-dual}

The eigenstates for the dynamical symmetries of the two algebraic
frameworks~--- number-conserving unitary and number-nonconserving
quasispin~--- are identical, that is, the irreps which reduce the
unitary algebra chains~(\ref{eqn-chainunbose})
and~(\ref{eqn-chainunfermi}) reduce the quasispin algebra
chains~(\ref{eqn-chainsu11}) and~(\ref{eqn-chainsu2}) as well, and the
labels for the chains are connected through the duality relations.
The pairing Hamiltonian is defined in Sec.~\ref{sec-model}
[see~(\ref{eqn-Hsum})] in terms of certain combinations of operators
which represent scalar pair creation, scalar pair annihilation, and
number operators.  These are noted in Sec.~\ref{sec-H-quasi}
[see~(\ref{eqn-H-quasi-gen})] to be essentially the quasispin
generators, and the
Hamiltonian can also therefore be expressed directly in terms of the
quasispin invariants [see~(\ref{eqn-H-quasi-invariant})].  However,
the pairing Hamiltonian can just as well be expressed in terms of the Casimir
operators of subalgebras of $\grpu{n_1+n_2}$, as a special case of the
Hamiltonian of Sec.~\ref{sec-unitary-dynsymm}
[see~(\ref{eqn-H-casimir})].  That such a relation exists is
implied by the duality of irreps, but it is explicitly obtained by appropriate
recoupling and reordering of the bosonic or fermionic creation and
annihilation operators in this section.

For a \textit{single} $j$-shell, recoupling and commutation of
creation operators\footnote{The product of a pair creation
operator and a pair annihilation operator is related to the product of
spherical-tensor one-body operators by, \textit{e.g.}, 
identity~(25a) of Ref.~\cite{chen1993:wick-coupled}.} yields
\begin{multline}
\label{eqn-recoupleaa}
(\ad\times\ad)^{(0)}_0(\at\times\at)^{(0)}_0
\nonumber\\
=
\frac{\theta}{\jhat^2}
\left[
-(\ad\times\at)^{(0)}_0+\sum_g\hat{g}[(\ad\times\at)^{(g)}\times(\ad\times\at)^{(g)}]^{(0)}_0
\right].
\end{multline}
Thus, the relation between quasispin and Casimir Hamiltonians for a
single level is 
\begin{equation}
\label{eqn-Casimirreln}
4S_+S_-=-\theta N +C_2[\grpu{n}]-\tfrac12\ctwogrpsosp{n},
\end{equation}
by comparison with the explicit realizations of the various
operators, namely, $N$ from Sec.~\ref{sec-unitary-casimir},
$C_2[\grpso{n}]$ or $C_2[\grpsp{n}]$ from~(\ref{eqn-C2sosp}), 
$C_2[\grpu{n}]$ from~(\ref{eqn-C2un}), and $S_\pm$ from~(\ref{eqn-Stensor}).

For the \textit{two-level} system, which has quasispin generators given
by~(\ref{eqn-S12}), the product $S_+S_-$ involves both single-level terms
($S_{1+}S_{1-}$ and $S_{2+}S_{2-}$) and ``cross terms'' ($\sigma
S_{1+}S_{2-}$ and $\sigma S_{2+}S_{1-}$), which destroy a
pair in one level and create a pair in the other.  Recoupling and
commutation of the mixed
terms yields
\begin{multline}
\label{eqn-recoupleab}
(\ad\times\ad)^{(0)}_0(\bt\times\bt)^{(0)}_0
+(\bd\times\bd)^{(0)}_0(\at\times\at)^{(0)}_0
\nonumber\\
=
\frac{(-)^{j_a+j_b}\theta\sigma_0}{\jhat_a\jhat_b}
\left[
\FcircF-2N_aN_b-\theta(\jhat_b^2N_a+\jhat_a^2N_b)
\right].
\end{multline}
The one-level terms and mixed terms of $S_+S_-$ may thus be combined to give an
expression involving the $\grpso{n_1+n_2}$ or $\grpsp{n_1+n_2}$
Casimir operator and $\grpu{n_1+n_2}$
invariants, {\it if and only if} the sign $\sigma$ arising in the definition
of the sum quasispin algebra and the sign $\sigma_0$ entering into
the definition of $F^{(g)}$ are related by 
\begin{equation}
\label{eqn-sigmacondition}
\frac{\sigma_0}{\sigma}=-\theta(-)^{j_a+j_b}.
\end{equation}
We again have an expression of the same
form as~(\ref{eqn-Casimirreln}), 
\begin{equation}
\label{eqn-Casimirrelnab}
4S_+S_-=-\theta (N_1+N_2) +C_2[\grpu{n_1+n_2}]-\tfrac12\ctwogrpsosp{n_1+n_2}.
\end{equation}
That the expression is of this form is to be expected from the general
nature of the duality, which indeed makes no assumption
(Sec.~\ref{sec-bose-fermi}) as to whether the single-particle states
are considered to be arranged into a single $j$-shell, as
for~(\ref{eqn-Casimirreln}), or two $j$-shells, as here.  From a practical
standpoint, what is most useful is that the relation can now be
expressed explicitly in terms of the two-level system operators given
by (\ref{eqn-C2sospab}), (\ref{eqn-C2unab}), and~(\ref{eqn-S12}) as
spherical-tensor products of creation and annihilation operators, with
well-defined phases $\sigma_0$ and $\sigma$.

The two-level operator correspondence~(\ref{eqn-Casimirrelnab}) relates the
$\grpcalsu[12]{1,1}$ or $\grpcalsu[12]{2}$ dynamical symmetry Hamiltonian in
the quasispin scheme~(\ref{eqn-H-quasi-equal}), \textit{i.e.}, strong
coupling with uniform pairing, to the $\grpso{n_1+n_2}$ or
$\grpsp{n_1+n_2}$ dynamical symmetry Hamiltonian in the unitary
algebra scheme.\footnote{The analog
of~(\ref{eqn-Casimirrelnab}) for the IBM was exploited in
Ref.~\cite{arima1979:ibm-o6} to establish the properties of the IBM
$\grpso{6}$ dynamical symmetry eigenstates.  With $\theta=+$, the sign
condition~(\ref{eqn-sigmacondition}) gives $\sigma_0=-\sigma$ (recall $j_a=0$,
and $j_b=2$). Thus,
the pairing operator $S_+S_-$ for the quasispin defined with {\it
negative} sign ($\sigma=-$) relates to the Casimir operator of the
``physical'' $\grpso{6}$ algebra ($\sigma_0=+$), and the quasispin
algebra defined with {\it positive} relative sign is instead dual to
the $\overline{\grpso{6}}$ algebra (Sec.~\ref{sec-unitary-gen}).}  
Furthermore, taken in conjunction with the single-level operator
correspondence~(\ref{eqn-Casimirreln}),
 it
allows the full two-level pairing Hamiltonian  (not just at the dynamical
symmetry limit) to be expressed in terms of Casimir operators of
algebras appearing in the two parallel subalgebra chains of
$\grpu{n_1+n_2}$.  Starting from~(\ref{eqn-H-quasi}), one obtains 
\begin{multline}
\label{eqn-H-un-invariant}
H=
(\varepsilon_1 -\tfrac14\theta G_{11})N_1
+\tfrac14(G_{11}-\sigma G_{12})C_2[\grpu[1]{n_1}]
-\tfrac18(G_{11}-\sigma G_{12})\ctwogrpsosp[1]{n_1}
\nonumber\\
\fl\qquad
+(\varepsilon_2 -\tfrac14\theta G_{22})N_2
+\tfrac14(G_{22}-\sigma G_{12})C_2[\grpu[2]{n_2}]
-\tfrac18(G_{22}-\sigma G_{12})\ctwogrpsosp[2]{n_2}
\nonumber\\
+\tfrac14\sigma G_{12} C_2[\grpu{n_1+n_2}]
-\tfrac18\sigma G_{12}\ctwogrpsosp{n_1+n_2}.
\end{multline}
For this relation to be valid, the phases $\sigma_0$ and $\sigma$,
used in defining the generators for orthogonal or symplectic algebra
and two-level quasispin algebra, respectively, must be related
by~(\ref{eqn-sigmacondition}).

\subsection{Multipole Hamiltonian}
\label{sec-multi}

The Hamiltonian for spectroscopic studies of the $s$-$b$ boson models
is commonly expressed in terms of a ``multipole'' term of the form
$[(\sd\times\bt)^{(L)}+(\bd\times\st)^{(L)}]
\cdot[(\sd\times\bt)^{(L)}+(\bd\times\st)^{(L)}]$, where $L=j_b$~\cite{vidal2006:lipkin-cqf-finite-size}.  For
instance, the customary IBM $\grpu{5}$--$\grpso{6}$ quadrupole
Hamiltonian~\cite{lipas1985:ibm-ecqf,iachello1998:phasecoexistence}
is
\begin{equation}
\label{eqn-Hqqu5so6}
H_{QQ}=\frac{(1-\xi)}{N} N_d -\frac{\xi}{N^2}
\bigl[(\sd\times\dt)^{(2)}+(\dd\times\st)^{(2)}\bigr]
\cdot\bigl[(\sd\times\dt)^{(2)}+(\dd\times\st)^{(2)}\bigr].
\end{equation}
The $\grpu{5}$ limit is obtained for $\xi=0$ and the $\grpso{6}$ limit
for $\xi=1$.  The operator $\FcircF$, appearing as the ``cross term''
in $C_2[\grpso{n_1+n_2}]$ or $C_2[\grpsp{n_1+n_2}]$
[see~(\ref{eqn-C2sospab})], generalizes the multipole term to the
generic two-level model.  In contrast, the Hamiltonian for
spectroscopic studies of the fermionic system is commonly expressed in
pairing or quasispin
form~\cite{vondelft2001:supercond-grain-spectroscopy,dusuel2005:bcs-scaling}.
Thus, we seek to relate these distinct~--- pairing and multipole~---
forms of the Hamiltonian.

Recall that the operators considered thus far in connection with the
strong-coupling limit~--- $\Svec^2$, $S_+S_-$, and
$C_2[\grpso{n_1+n_2}]$ or $C_2[\grpsp{n_1+n_2}]$~--- differ only in
normalization (or sign) and by addition of a function of $N$, the
conserved total occupation number.  Therefore, the eigenstates are
identical and the eigenvalues differ only by a rescaling and a
constant offset.  However, the operator $\FcircF$ appearing in the
multipole Hamiltonian differs from these 
[again, see~(\ref{eqn-C2sospab})] by terms proportional to
$C_2[\grpso[1]{n_1}]$ and $C_2[\grpso[2]{n_2}]$, in the bosonic case,
or $C_2[\grpsp[1]{n_1}]$ and $C_2[\grpsp[2]{n_2}]$, in the fermionic
case.  The \textit{eigenstates} are therefore again the same as for
$\Svec^2$, $S_+S_-$, and $C_2[\grpso{n_1+n_2}]$ or
$C_2[\grpsp{n_1+n_2}]$, but \textit{eigenvalues} are no longer
degenerate for states sharing the same value of $v$.  They are rather now
split by $v_1$ and $v_2$, as illustrated in Fig.~\ref{fig-schemes}.

For the explicit relationships among these operators, observe that,
by~(\ref{eqn-Casimirrelnab}),
\begin{equation}
\label{eqn-Casimirrelnab-SS-C}
4S_+S_-= C_2[\grpu{n_1+n_2}]-\theta N-\tfrac12\ctwogrpsosp{n_1+n_2},
\end{equation}
and, in terms of~(\ref{eqn-C2sospab}),
\begin{equation}
\label{eqn-Casimirrelnab-C-FF}
\ctwogrpsosp{n_1+n_2}=2\theta\FcircF+ \ctwogrpsosp[1]{n_1} +\ctwogrpsosp[2]{n_2}.
\end{equation}
Therefore, the pairing ($S_+S_-$) and multipole ($\FcircF$) forms of
the Hamiltonian, those most
frequently encountered in applications, are related by
\begin{multline}
\label{eqn-Casimirrelnab-FF-SS}
-\theta\FcircF=4S_+S_- 
-\Bigl[C_2[\grpu{n_1+n_2}]-\theta N\Bigr]
\nonumber\\
+\frac12 \ctwogrpsosp[1]{n_1}
+\frac12 \ctwogrpsosp[2]{n_2},
\end{multline}
where 
\begin{equation}
\label{eqn-C2diff-eigen}
C_2[\grpu{n_1+n_2}]-\theta N
=
\Biggl\lbrace\begin{gathered}
N(N+2\Omega-2)
\\
N(2\Omega-N+2)
\end{gathered}
\end{equation}
simply contributes a $c$-number shift to the eigenvalue spectrum,
without affecting the eigenfunctions.  

A positive coefficient ($G>0$) for $S_+S_-$ gives a positive pair energy,
\textit{i.e.}, repulsive pairing, in both bosonic and fermionic cases,
as may be seen from~(\ref{eqn-SS-eigen}) with $N=1$ and $v=0$. The sign of the pairing
interaction is of special interest in comparing the bosonic and
fermionic two-level pairing models, since it should be noted
(Sec.~\ref{sec-trans}) that the system undergoes a quantum phase
transition for {\it repulsive} ($G>0$) pairing interaction in the
bosonic case and {\it attractive} pairing interaction ($G<0$) in the
fermionic case.  Thus, it is essential to note that repulsive pairing
is obtained for a negative coefficient on $\FcircF$ in the bosonic
case and a positive coefficient on $\FcircF$ in the fermionic case,
{\it i.e.}, for $H=-\theta\FcircF$, or \textit{vice versa} for attractive
pairing.  Therefore, for repulsive pairing, a Hamiltonian 
\begin{equation}
\label{eqn-H-FF}
H_{FF}=\frac{(1-\xi)}{N}N_2 - \theta
\frac{\xi}{N^2}\FcircF
\end{equation}
is the natural generalization of the multipole form for the
transitional Hamiltonian~(\ref{eqn-Hqqu5so6}) to generic two-level
pairing models, as considered in Sec.~\ref{sec-trans}.

The last two in terms in~(\ref{eqn-Casimirrelnab-FF-SS}), involving
the Casimir operators of $\grpso[1]{n_1}$ and $\grpso[2]{n_2}$ or
$\grpsp[1]{n_1}$ and $\grpsp[2]{n_2}$, contribute a common shift to
the energy eigenvalues for each subspace of states characterized by
a given pair of values of the conserved $(v_1v_2)$ quantum numbers, without
affecting the eigenfunctions, \textit{i.e.}, these terms serve only to displace the
different $(v_1v_2)$ subspaces relative to each other.  If only
$(v_1v_2)=(00)$ states are considered, the $\grpso[1]{n_1}$ and
$\grpso[2]{n_2}$ or $\grpsp[1]{n_1}$ and $\grpsp[2]{n_2}$ terms have
no effect at all.  They will therefore not be considered further.

\begin{figure}
\begin{center}
\includegraphics*[width=0.8\hsize]{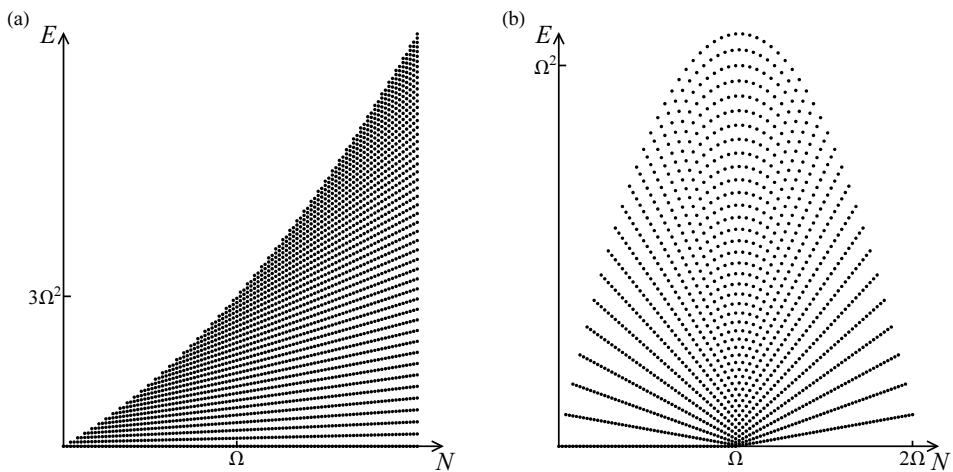}
\end{center}
\caption{Eigenvalues of the pairing interaction term $4S_+S_-$, which
determines the energy spectrum of the two-level pairing model in the
strong coupling limit, shown for the (a)~bosonic and (b)~fermionic
systems, as a function of filling $N$.  The eigenvalues are given
by~(\ref{eqn-SS-eigen-range-bose})
and~(\ref{eqn-SS-eigen-range-fermi}), respectively.  The axes are
labeled generically, to indicate the asymptotic (large-$\Omega$)
dependences discussed in the text, but the specific points shown for
illustration are calculated for $\Omega=50$.}
\label{fig-filling}
\end{figure}

Now to consider the spectra, the strong-coupling Hamiltonian operator
$4S_+S_-$~--- we include the factor of $4$ arising
in~(\ref{eqn-Casimirrelnab-FF-SS}) for convenience~--- has eigenvalues
given by~(\ref{eqn-SS-eigen}), obtained with $0\leq v\leq N$ for
bosonic pairing or with $0\leq v \leq \min(N,2\Omega-N)$ for fermionic
pairing, where only even values of $v$ arise for $N$ even, or odd
values of $v$ for $N$ odd [see~(\ref{eqn-branchuso})
and~(\ref{eqn-branchusp})].  Thus, taking $N$ even, the eigenvalues
span the range
\begin{equation}
\label{eqn-SS-eigen-range-bose}
\langle4S_+S_-\rangle=\underbrace{0}_{v=N},\ldots,
\underbrace{N(N+2\Omega-2)}_{v=0},
\end{equation}
for bosonic pairing, or
\begin{equation}
\label{eqn-SS-eigen-range-fermi}
\langle4S_+S_-\rangle=\underbrace{0}_{\substack{v=N\\(N\leq\Omega)}}~\text{or}~\underbrace{4(N-\Omega)}_{\substack{v=2\Omega-N\\(N\geq\Omega)}},
\ldots,
\underbrace{N(2\Omega-N+2)}_{v=0},
\end{equation}
for fermionic pairing, in which case $0\leq N\leq 2\Omega$.  (If $N$
is odd, the sequences above would end instead with $v=1$ rather than
$v=0$, but the large-$N$ dependence of the highest eigenvalue on $N^2$
is not changed.)  The range of eigenvalues therefore depends upon the
total occupation or ``filling'' $N$ of the two-level system as
sketched in Fig.~\ref{fig-filling}(a) for the bosonic system and
Fig.~\ref{fig-filling}(b) for the fermionic system.  The asymptotic
dependences for large degeneracy ($\Omega\gg 1$) are indicated.
Specifically, at a
filling approximately equal to half the total degeneracy,
\textit{i.e.}, $N\approx\Omega$, note that the bosonic eigenvalues
span a range $\sim 3\Omega^2(\approx3N^2)$, while the fermionic
eigenvalues for the same filling and degeneracy only span a range of
$\sim \Omega^2(\approx N^2)$.  If, instead, the limit of large
occupation is taken at fixed degeneracy ($\Omega\ll N$) in the bosonic
case, the range of eigenvalues $\sim N^2$ is the same as for a
fermionic pairing model of the same $N$ but at half filling (which is
obtained for a correspondingly larger degeneracy $\Omega=N$).
%--------------------------------
%--------------------------------

Note also that, for repulsive pairing, the bosonic ground state (for
which $v=N$) has zero eigenvalue.  In contrast, the fermionic ground
state (for which $v=0$ below half filling and $v=2\Omega-N$ past half
filling) has an eigenvalue which grows linearly with $N$ past half
filling.
%%\footnote{
The nonzero ground state pairing energy for the
fermionic system  may be understood since, past half filling, Pauli
exclusion enforces the existence of some particles in time-reversal
conjugate orbits and hence some probability for pairs coupled to zero
angular momentum.
%%}

For the multipole form
$-\theta\FcircF$ of the pairing interaction operator, the spectrum is
shifted downward by an $N$-dependent offset relative to that of $4S_+S_-$ [see~(\ref{eqn-Casimirrelnab-FF-SS})].  The {\it
highest} eigenvalue (obtained for $v=0$) is always zero.  The
asymptotic form of the ground state eigenvalue is, alternatively,
$\sim -N^2$ for fermionic half filling ($1\ll \Omega = N$), $\sim
-3N^2$ for bosonic ``half filling'' ($1\ll \Omega \approx N$), and
again $\sim -N^2$ for the bosonic system at larger boson number
($\Omega \ll N$).  

\section{Transitional Hamiltonian}
\label{sec-trans}

A second-order quantum
state phase transition occurs between the weak-coupling and
strong-coupling limits for the two-level pairing models.
Specificially, for the {\it bosonic} system it occurs with {\it
repulsive} pairing interaction, and  for the
{\it fermionic} system it occurs with {\it attractive} interaction.
The quantum phase transition is apparent numerically from calculations
for finite $N$ and from semiclassical treatments of the large-$N$
limit.  The present duality relations (Sec.~\ref{sec-hamiltonian})
immediately help clarify comparison of numerical eigenvalue spectra
across the transition but are also intended to facilitate the
construction of coherent states for the semiclassical treatment.  

The simplest semiclassical ``geometry'' for the two-level pairing
model is obtained from the quasispin algebraic structure, most simply
by replacing the quasispin operators with classical angular momentum
vectors, which maps the pairing model onto an essentially
one-dimensional coordinate space.  This approach has been applied in
both the bosonic $s$-$b$ models and fermionic two-level pairing model
with equal
degeneracies~\cite{feng1981:ibm-phase,somma2004:qpt-entanglement,dusuel2005:bcs-scaling,leyvraz2005:lipkin-scaling,tsue2007:two-level-semiclassical}.
In both these circumstances, the quantum phase transition is found to
occur, in the large-$N$ limit, at $N\abs{G}/\varepsilon=1$.  For the
$s$-$b$ models, a higher-dimensional and richer classical geometry
(see Ref.~\cite{feng1981:ibm-phase}) has been established through the
use of $\grpu{n_2+1}/\grpu{n_2}$ coherent
states~\cite{gilmore1978:lipkin,feng1981:ibm-phase,dieperink1980:ibm-classical,ginocchio1980:ibm-coherent-bohr,cejnar2007:phase-cusp-un}.
An extension of this treatment to
$\grpu{n_1+n_2}/[\grpu{n_1}\otimes\grpu{n_2}]$ for generic two-level
pairing models might profitably be obtained using the explicit
construction of generators for the
$\grpso{n_1+n_2}\supset\grpso{n_1}\otimes\grpso{n_2}$ and
$\grpsp{n_1+n_2}\supset\grpsp{n_1}\otimes\grpsp{n_2}$ chains
considered in Sec.~\ref{sec-unitary}.

\begin{figure}
\begin{center}
\includegraphics*[width=0.5\hsize]{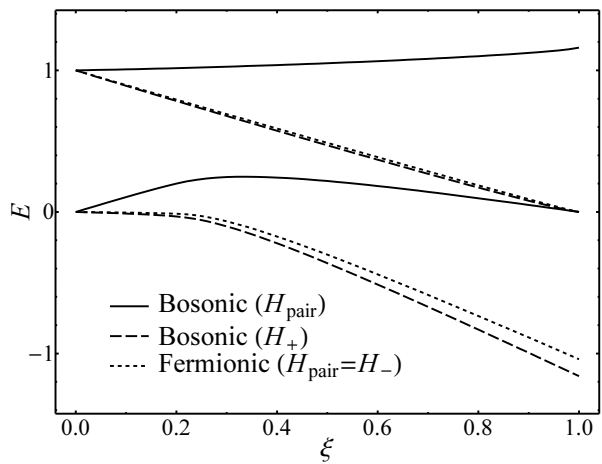}
\end{center}
\caption{Lowest and highest eigenvalues, within the $(v_1v_2)=(00)$ subspace, for the bosonic two-level
pairing model with repulsive pairing interaction, in the usual form as
given by $H_{\text{pair}}$ in~(\ref{eqn-H-trans-pair}) (solid curves)
and after correction by the $\grpu{n_1+n_2}$ Casimir offset as given
by $H_+$ in~(\ref{eqn-H-trans-plus}) (dashed curves).  These are
plotted as functions of the control parameter $\xi$ between the
weak-coupling and strong-coupling limits.  The eigenvalues for the
fermionic two-level pairing model with attractive pairing interaction,
as given by $H_{\text{pair}}$ or equivalently by $H_-$
in~(\ref{eqn-H-trans-minus}), are shown for comparison (dotted
curves).  All calculations are for $N=50$, with $n_1=n_2=5$ in the
bosonic case and $n_1=n_2=50$ in the fermionic case (see
Fig.~\ref{fig-correlation-N50} for more detailed eigenvalue spectra).
}
\label{fig-envelope-N50}
\end{figure}

However, at present, we confine ourselves to laying the groundwork for
more detailed further work, allowing for the most general choice of
level degeneracies and more uniformly treating the bosonic and
fermionic cases.  A pairing Hamiltonian
\begin{equation}
\label{eqn-H-trans-pair}
H_{\text{pair}}=\frac{(1-\xi)}{N}N_2
+\theta \frac{4\xi}{N^2}S_+S_-,
\end{equation}
may be defined with opposite signs $\theta$ of the pairing term for
the bosonic and fermionic cases, so that the quantum phase transition
is obtained in either case.  This Hamiltonian yields the weak-coupling
limit for $\xi=0$, the strong-coupling limit at $\xi=1$, and the
critical interaction strength $N\abs{G}/\varepsilon=1$ at $\xi=1/5$.
Scaling of the one-body term by $N^{-1}$ and of the two-body term by
by $N^{-2}$ ensures that the critical point remains fixed at the
finite value $\xi=1/5$ as $N\rightarrow\infty$.  However, for this
Hamiltonian, a grossly different ``envelope'' to the eigenvalue
spectrum (\textit{i.e.}, the range of eigenvalues, 
obtained as a function of the control parameter $\xi$) is found in the
bosonic and fermionic cases (see Fig.~\ref{fig-envelope-N50}).
%--------------------------------
%--------------------------------

To facilitate direct
comparison of the bosonic and fermionic quantum phase transitions, it
is helpful to instead construct a Hamiltonian for which the ground
state energy follows the same trajectory as a function of $\xi$ in the
large $N$ limit, and the eigenvalues span the same range at each of
the limits, namely, $[0,1]$ for $\xi=0$ and $[-1,0]$ for $\xi=1$.  By
the results of Sec.~\ref{sec-hamiltonian}, this is accomplished by
choosing, for {\it repulsive} pairing interaction, the Hamiltonian
\begin{equation}
\label{eqn-H-trans-plus}
H_+=\frac{(1-\xi)}{N}N_2 + \frac{\xi}{N^2}\Biggl[4S_+S_-
-\Biggl\lbrace\begin{gathered}
N(N+2\Omega-2)
\\
N(2\Omega-N+2)
\end{gathered}
\Biggr\rbrace
\Biggr],
\end{equation}
and, for {\it attractive} pairing interaction, the usual Hamiltonian
\begin{equation}
\label{eqn-H-trans-minus}
H_-=\frac{(1-\xi)}{N}N_2 - \frac{4\xi}{N^2}S_+S_-.
\end{equation}
The $c$-number offset included in the definition of $H_+$, which
arises as $C_2[\grpu{n_1+n_2}]-\theta N$, is included to achieve the
same range of eigenvalues in the strong coupling limit, as well as a
similar evolution of ground state energy across the transition
(Fig.~\ref{fig-envelope-N50}), in the large-$N$ limit, thereby
facilitating comparison of the bosonic (repulsive pairing) and
fermionic (attractive pairing) quantum phase transition.  With
inclusion of this offset, $H_+$ is equivalent to the generalized
multipole transitional Hamiltonian~(\ref{eqn-H-FF}) when acting on the
$(v_1v_2)=(00)$ subspace of any two-level pairing model, as may be
seen from~(\ref{eqn-Casimirrelnab-FF-SS}).  In particular, for the
$s$-$b$ models, inclusion of this offset makes $H_+$ identical to the
conventional multipole form~(\ref{eqn-Hqqu5so6}) of the transitional
Hamiltonian, when acting on the $(v_1v_2)=(00)$ subspace.

\begin{figure}
\begin{center}
\includegraphics*[width=0.8\hsize]{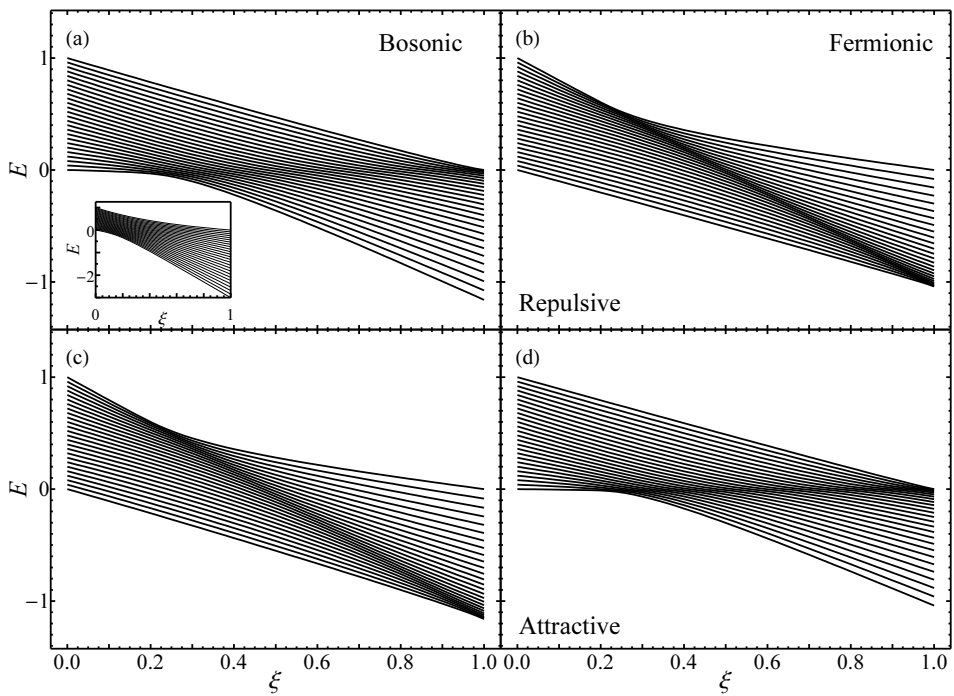}
\end{center}
\caption{Eigenvalues of the bosonic two-level pairing model, with
level degeneracies $n_1=n_2=5$ (at left), and fermionic two-level
pairing model, with level degeneracies $n_1=n_2=50$ (at right), for
repulsive (at top) and attractive (at bottom) pairing interactions,
shown for the $(v_1v_2)=(00)$ subspace, as functions of the control
parameter $\xi$ between the weak-coupling and strong-coupling limits.
All calculations are for $N=50$, thus for $\Omega\ll N$ in the bosonic
case and $\Omega=N$ (half filling) in the fermionic case.  The
alternative regime in which the bosonic system also has $\Omega \approx
N$ is shown (specifically, for $n_1=n_2=51$ and $N=50$) in the inset
to panel~(a).  The Hamiltonians $H_\pm$ of~(\ref{eqn-H-trans-plus})
and~(\ref{eqn-H-trans-minus}) are used in the calculations.
}
\label{fig-correlation-N50}
\end{figure}

The evolution of the eigenvalue spectrum across the transition
between weak coupling and strong coupling is shown for representative
bosonic and fermionic cases in
Fig.~\ref{fig-correlation-N50}.  Specifically, equal-degeneracy
pairing models ($n_1=n_2\equiv\Omega$) are considered, and the
$(v_1v_2)=(00)$ states are shown.  Here
a sufficiently large total occupancy ($N=50$) is chosen such that the
precursors of the phase transitional singularities are readily
apparent.  Spectra are shown for both bosons
[Fig.~\ref{fig-correlation-N50}~(left)] and fermions
[Fig.~\ref{fig-correlation-N50}~(right)], with repulsive
[Fig.~\ref{fig-correlation-N50}~(top)] and attractive
[Fig.~\ref{fig-correlation-N50}~(bottom)] interactions.  Qualitatively
similar spectra in the bosonic and fermionic cases are obtained when level
degeneracies for the bosonic calculation ($\Omega=5$) are \textit{much less
than} the occupancy, while the
degeneracies for the fermionic calculation ($\Omega=50$) are such as
to give \textit{half filling}.  Then the ``envelope'' of the spectrum
(the range of eigenvalues at a given value of the Hamiltonian parameter
$\xi$) is essentially identical for the \textit{bosonic} case with \textit{repulsive}
pairing [Fig.~\ref{fig-correlation-N50}~(a)] and the \textit{fermionic} case
with \textit{attractive} pairing [Fig.~\ref{fig-correlation-N50}~(d)], {\it
i.e.}, the interactions signs which yield a ground state quantum phase
transition.  Features to observe include the essentially constant
ground state energy for $\xi<1/5$ and downturn [from $0$ to $-1$] for
$\xi>1/5$, an approximately linear evolution of the highest eigenvalue
from $+1$ to $0$, and a compression of the level density at $E\approx
0$ for $\xi>1/5$, a characteristic of the excited state quantum phase
transition~\cite{caprio2008:esqpt}.
%--------------------------------
%--------------------------------

The structure of the eigenvalue spectrum is likewise
similar
when one compares the \textit{bosonic} case with \textit{attractive} pairing
[Fig.~\ref{fig-correlation-N50}~(c)] and the \textit{fermionic} case with
\textit{repulsive} pairing [Fig.~\ref{fig-correlation-N50}~(b)].
This should hardly be surprising.  Indeed, when
$n_1=n_2$, the eigenvalue spectra for Hamiltonians for opposite pairing signs [\textit{e.g.},
Fig.~\ref{fig-correlation-N50}~(a) and
Fig.~\ref{fig-correlation-N50}~(c), or
Fig.~\ref{fig-correlation-N50}~(b) and Fig.~\ref{fig-correlation-N50}~(d)]
may be obtained from each other, by negation of the Hamiltonian and interchange of the level
labels 1 and~2, to within addition of a $c$-number function of $\xi$.
Therefore, in the present example, the resemblance between Fig.~\ref{fig-correlation-N50}~(c) and
Fig.~\ref{fig-correlation-N50}~(b) is a necessary consequence of the resemblance
between Fig.~\ref{fig-correlation-N50}~(a) and
Fig.~\ref{fig-correlation-N50}~(d).

The emergence of finite-size precursors to the infinite-$N$
singularities associated with the quantum phase transition depends not
only on $N$ but also on the level degeneracies $n_1$ and $n_2$.  An
important distinction therefore arises between bosonic and fermionic
models~\cite{caprio2008:esqpt}.
For {\it fermionic} systems, the
total occupancy $N$ is limited to $n_1+n_2$.  Therefore, the limit of
large $N$ can only be taken if the level degeneracies are
simultaneously increased.  Since at full filling ($N=n_1+n_2$) the
spectrum, like that for zero filling, is trivial, it is more
informative to take the limit $N\rightarrow\infty$ at or near half filling
[$N=\tfrac12(n_1+n_2)\equiv\Omega$].
However, no such restriction arises for {\it bosonic} systems, and
$N\rightarrow\infty$ can be obtained even for fixed level
degeneracies.

Indeed, for the bosonic two-level pairing models, we find numerically
that the onset of critical phenomena requires $N\gg\Omega$, not
$N\approx\Omega$.  The evolution of eigenvalues for the bosonic system
with the same occupancy ($N=50$) as in
Fig.~\ref{fig-correlation-N50}(a), but with level degeneracies
comparable to the occupation $n_1=n_2(=\Omega)=49\approx50$, analogous
to ``half filling'', is shown for comparison in
Fig.~\ref{fig-correlation-N50}~(inset).  The eigenvalue spectrum is
qualitatively different, as compared to
Fig.~\ref{fig-correlation-N50}(a) or~(d), with respect to each of the
properties noted above, {\it e.g.}, the ground state eigenvalue is not
recognizably constant for $\xi<1/5$, there is no apparent change in
curvature at $\xi=1/5$, and closer inspection reveals no level spacing
compression of the type associated with the excited state quantum
phase transition.  This is already anticipated from the different
eigenvalue range ($\sim3N^2$) in the strong coupling limit, obtained
in Sec.~\ref{sec-multi}.

Similar distinctions between the large-$N$ limit taken with $N \gg
\Omega$ or $N \sim \Omega$ are obtained for the critical scaling
properties, which we defer to a more comprehensive study.  For now, we
restrict attention to the basic energy spectra obtained with the
present transitional Hamiltonian for the general two-level pairing
model, and note that the spectrum for finite $N$ depends strongly not
just on the total degeneracy $n=n_1+n_2$ of the two levels but on the
equality or degree of inequality of the two level degeneracies $n_1$
and $n_2$.  
%--------------------------------
\begin{figure}
\begin{center}
\includegraphics*[width=0.7\hsize]{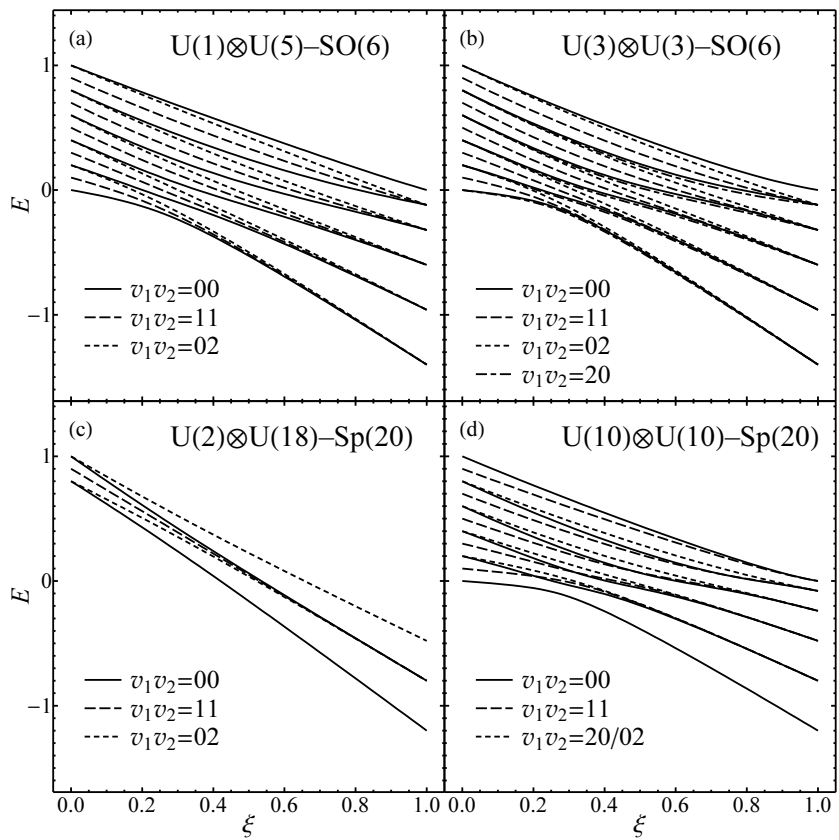}
\end{center}
\caption{  
Eigenvalues for the bosonic two-level pairing model, with level
degeneracies (a)~$n_1=1$ and $n_2=5$ or (b)~$n_1=3$ and $n_2=3$, and
for the fermionic two-level pairing model, with level degeneracies
(c)~$n_1=2$ and $n_2=18$ or (d)~$n_1=10$ and $n_2=10$, as functions of
the control parameter $\xi$ between the weak-coupling and
strong-coupling limits.  All calculations are for $N=10$.  Eigenvalues
are shown only for the lowest-seniority subspaces $(v_1v_2)$,
specifically, those with $v_1+v_2\leq 2$.  The Hamiltonian $H_+$
of~(\ref{eqn-H-trans-plus}) is used for the bosonic calculations and
$H_-$ of~(\ref{eqn-H-trans-minus}) for the fermionic calculations.}
\label{fig-correlation-N10}
\end{figure}
%--------------------------------

The transitional spectra for two different bosonic models
with total degeneracy $n=6$, and taken with $N=10$ (\textit{i.e.},
occupation substantially greater than the degeneracy), are compared in
Fig.~\ref{fig-correlation-N10}~(top): the $s$-$b$ model ($n_1=1$ and
$n_2=5$) [Fig.~\ref{fig-correlation-N10}(a)] and the choice of two
levels with equal
degeneracies ($n_1=3$ and
$n_2=3$)  [Fig.~\ref{fig-correlation-N10}(b)\footnote{Fig.~\ref{fig-correlation-N10}(b) also corrects a
labeling error in the legend of Fig.~6(c) of
Ref.~\cite{caprio2008:esqpt}.}].
The transitional spectra for the $(v_1v_2)=(00)$ subspaces in
Figs.~\ref{fig-correlation-N10}(a) and~(b) are similar to each other.
Although only irreps of type $(0v_2)$ or $(1v_2)$ are obtained in the
former case, more general irreps $(v_1v_2)$ are possible in the latter
case, naturally leading to a more complicated spectrum.  In particular, it
should be noted that the lowest state from each subspace of the form
$(v_10)$ approximately tracks the lowest $(00)$ state in energy, and
that these states are in fact lower in energy than the $(00)$ state everywhere between the
dynamical symmetry limits [see the lowest curve for $(v_1v_2)=(20)$ in
Fig.~\ref{fig-correlation-N10}(b)].    It is perhaps not surprising
that, given a repulsive pairing interaction, the energy may be lowered
by breaking pairs within the lower single-particle energy
(\textit{i.e.}, increasing $v_1$).  In contrast, increasing $v_2$ also
enforces nonzero occupation ($N_2\geq v_2$) of the higher
single-particle energy level and is therefore not as energetically prefered.

For the fermionic system, the difference
between the transitional spectra for near-equal versus
highly-imbalanced degeneracies for the two levels is marked.  The
transitional spectra for two different fermionic models with total
degeneracy $n=20$, again taken with $N=10$ (which now represents half
filling), are compared in Fig.~\ref{fig-correlation-N10}~(bottom): for
the most extremely imbalanced possible choice of degeneracies ($n_1=2$
and $n_2=18$) [Fig.~\ref{fig-correlation-N10}(c)] and with equal
degeneracies ($n_1=10$ and $n_2=10$)
[Fig.~\ref{fig-correlation-N10}(d)].  The quantum phase transition
which occurs for equal degeneracies is washed out in the limit of
imbalanced degeneracies, as is evident in the simple, near-linear
evolution of the ground state energy across
Fig.~\ref{fig-correlation-N10}(c).  Such an effect may be expected on
the basis of the Pauli principle.  The lower level, of degeneracy
$n_1=2$, easily saturates at full occupancy, so that the dynamics are
effectively those of a one-level system of degeneracy $n_2=18$, which
does not support critical phenomena as a function of pairing
interaction strength.

\section{Conclusion}
\label{sec-concl}

Although the existence of duality relations between the
number-conserving unitary and number-nonconserving quasispin algebras
for the two-level system with pairing interactions is well known, and
indeed these relations have proven useful in practical calculations
for specific special cases of the two-level pairing model, here we
have sought to establish a systematic treatment of the duality
relations, both for bosonic and fermionic two-level pairing models and
for arbitrary choice of level degeneracies.  A principal goal has been
to clarify the relationships between the disparate forms of the
Hamiltonian encountered in the study and application of these models.
The results are intended to provide a foundation for a more
comprehensive investigation of quantum phase transitions in two-level
pairing models~--- including the dependence of scaling properties on
the bosonic or fermionic nature of the system and on the level
degeneracies~--- beyond the special cases conventionally considered,
namely, bosonic $s$-$b$ models and fermionic models with equal
degeneracy.  The duality between orthogonal or symplectic algebras and
the quasispin algebras is also relevant to the analysis of the
classical dynamics of the system, through the associated coset
spaces~\cite{gilmore1974:lie-groups}.  The dual algebras yield
complementary descriptions involving classical coordinate spaces with
different dimensionalities~\cite{feng1981:ibm-phase}.  Finally,
although the present derivations were given for the case of two-level
models, they may readily be extended to the $\grpso{n_1+n_2+\cdots}$
or $\grpsp{n_1+n_2+\cdots}$ algebras associated with multi-level
systems, with generators directly generalizing those
of~(\ref{eqn-Gkkdefn}) and~(\ref{eqn-Feta}), for which the quantum
phase transitions have been much less completely studied.  Physical
realizations of interest in this more general case include the
nuclear shell model and descriptions of superconductivity in metallic grains.

%\begin{acknowledgments}
\ack
We thank V.~Hellemans for comments on the manuscript.  This work was
supported by the US DOE under grants DE-FG02-95ER-40934 and
DE-FG02-91ER-40608.
%%\end{acknowledgments}

\appendix
\def\thesection{Appendix}   
\section{Spherical tensor commutation relations}
\label{sec-appcoupled}

When
working with angular momentum coupled products of spherical tensor
operators, it is convenient to consider the {\it coupled
commutator}~\cite{french1966:multipole,chen1993:wick-coupled}, itself
a spherical tensor operator, with components given by
\begin{equation}
\label{eqn-comm-coupled-components}
[A^{(a)},B^{(b)}]^{(c)}_\gamma=\sum_{\alpha\beta}\tcg{a}{\alpha}{b}{\beta}{c}{\gamma}
[A^{(a)}_\alpha,B^{(b)}_\beta].
\end{equation}
To clearly set out the identities used in establishing the commutators of
the generators in Tables~\ref{tab-comm-un} and~\ref{tab-comm-sosp},
the basic
definitions and properties are summarized in this appendix.  
The use of coupled commutation results bypasses the tedious
process of uncoupling the operators, \textit{i.e.}, introducing multiple sums over
products of Clebsch-Gordan coefficients, taking commutators of the
spherical tensor components, and then recoupling.

If both bosonic and fermionic operators are to be considered, a
consistent set of definitions is obtained if the quantity in brackets
is taken to be the {\it graded} commutator, that is, either the
commutator or the anticommutator according to the bosonic or fermionic
nature of the operators.  Specifically,
\begin{equation}
\label{eqn-comm-graded}
[A^{(a)}_\alpha,B^{(b)}_\beta]=A^{(a)}_\alpha B^{(b)}_\beta
-\theta_{ab} B^{(b)}_\beta A^{(a)}_\alpha,
\end{equation}
where $\theta_{ab}=+$ if either $A$ or $B$ is a bosonic operator, and
$\theta_{ab}=-$ if both $A$ and $B$ are fermionic operators.  (For the
sake of these definitions, it is assumed that a bosonic operator has
integer angular momentum and a fermionic operator has half-integer angular momentum.)
The coupled commutator can be written directly in terms of coupled
products as
\begin{equation}
\label{eqn-comm-coupled-prod}
[A^{(a)},B^{(b)}]^{(c)}
=
(A^{(a)}\times B^{(b)})^{(c)}
-\theta_{ab}(-)^{c-a-b}
(B^{(b)}\times A^{(a)})^{(c)}
\end{equation}
and obeys the symmetry or antisymmetry relation
\begin{equation}
\label{eqn-comm-symm}
[B^{(b)},A^{(a)}]^{(c)}
=
-\theta_{ab}(-)^{c-a-b}
[A^{(a)},B^{(b)}]^{(c)}.
\end{equation}
The uncoupled commutators of the spherical tensor components
may be recovered from the coupled commutators, if needed, by
inverting~(\ref{eqn-comm-coupled-components}) to give
\begin{equation}
\label{eqn-comm-uncoupled}
[A^{(a)}_\alpha,B^{(b)}_\beta]
=
\sum_{c\gamma}
\tcg{a}{\alpha}{b}{\beta}{c}{\gamma}
[A^{(a)},B^{(b)}]^{(c)}_\gamma.
\end{equation}

The product rule for coupled commutators
is~\cite[(6)]{chen1993:wick-coupled}
\begin{multline}
\label{eqn-coupled-product-rule}
[(A\times B)^{(e)},C]^{(d)}
=\sum_f
\hat{e}\hat{f}
\bigl[
(-)^{a+b+c+d}
\smallsixj{a}{b}{e}{c}{d}{f}
(A\times[B,C]^{(f)})^{(d)}
\nonumber
\\+
\theta_{bc}
\smallsixj{a}{b}{e}{d}{c}{f}
([A,C]^{(f)}\times B)^{(d)}
\bigr].
\end{multline}
A second application of this identity yields the double product rule
needed for evaluating commutators of one-body or pair
operators,
\begin{multline}
\label{eqn-coupled-double-product-rule}
[(A\times B)^{(e)},(C\times D)^{(f)}]^{(g)}
\nonumber\\\fl\qquad
=\sum_{hk}
\hat{e}\hat{f}\hat{h}\hat{k}
\bigl[
(-)^{e+c+d+g}(-)^{a+b+c+h}
\smallsixj{c}{d}{f}{g}{e}{h}
\smallsixj{a}{b}{e}{c}{h}{k}
[(A\times[B,C]^{(k)})^{(h)}\times D]^{(g)}
\nonumber\\\fl\qquad
+
\theta_{bc}
(-)^{e+c+d+g}(-)^{b+c+e+k}
\smallsixj{c}{d}{f}{g}{e}{h}
\smallsixj{a}{b}{e}{h}{c}{k}
[([A,C]^{(k)}\times B)^{(h)}\times D]^{(g)}
\nonumber\\\fl\qquad
+
\theta_{ec}
(-)^{e+c+f+h}(-)^{a+b+d+h}
\smallsixj{c}{d}{f}{e}{g}{h}
\smallsixj{a}{b}{e}{d}{h}{k}
[C\times (A\times [B,D]^{(k)})^{(h)}]^{(g)}
\nonumber\\\fl\qquad
+
\theta_{ec}\theta_{bd}
(-)^{e+c+f+h}(-)^{b+d+e+k}
\smallsixj{c}{d}{f}{e}{g}{h}
\smallsixj{a}{b}{e}{h}{d}{k}
[C\times([A,D]^{(k)}\times B)^{(h)}]^{(g)}
\bigr]
.
\end{multline}
If operators $A^\dagger$ and $B^\dagger$ are creation operators,
obeying cannonical commutation or anticommutation relations, the
canonical commutators are represented in coupled form
by~\cite[(10)]{chen1993:wick-coupled}
\begin{equation}
\label{eqn-comm-canonical}
[\tilde{A},B^\dagger]^{(c)}=\hat{a}\delta_{AB}\delta_{c0}
\end{equation}
and $[\tilde{A},\tilde{B}]^{(c)}=[A^\dagger,B^\dagger]^{(c)}=0$.  The
coupled commutator of two one-body operators is therefore
\begin{multline}
\label{eqn-comm-bilinear}
[(A^\dagger \times \tilde{B})^{(e)},(C^\dagger \times \tilde{D})^{(f)}]^{(g)}
=
(-)^{2b}(-)^{a+d+g}
\hat{e}\hat{f}
\smallsixj{e}{f}{g}{d}{a}{b}
(A^\dagger\times\tilde{D})^{(g)}
\delta_{BC}
\nonumber\\
-
\theta_{ab}\theta_{bc}\theta_{ca} (-)^{b+c+e+f}
\hat{e}\hat{f}
\smallsixj{e}{f}{g}{c}{b}{a}
(C^\dagger\times\tilde{B})^{(g)}
\delta_{AD},
\end{multline}
as needed, \textit{e.g.}, for the commutators of the generators of
$\grpu{n_1+n_2}$.

%***************************************************************************
% bibliography
%***************************************************************************

\section*{References}
\input{pairalg.bbl}%bibliography{master,mc,theory,expt,books,pairalg,misc}

\end{document}

%% file: pairalg_tab01.tex
\begin{ruledtabular}
\begin{tabular}{rrl}
\br
$v$ & $n_v$ & $(v_1,v_2)$ \\
\mr
0&0&(0,0)\\
1&0&(1,0), (0,1)\\
2&0&(2,0), (1,1), (0,2)\\
 &1&(0,0)\\
3&0&(3,0), (2,1), (1,2), (0,3)\\
 &1&(1,0), (0,1)\\
4&0&(4,0), (3,1), (2,2), (1,3), (0,4)\\
 &1&(2,0), (1,1), (0,2)\\
 &2&(0,0)\\
5&0&(5,0), (4,1), (3,2), (2,3), (1,4), (0,5)\\
 &1&(3,0), (2,1), (1,2), (0,3)\\
 &2&(1,0), (0,1)\\
6&0&(6,0), (5,1), (4,2), (3,3), (2,4), (1,5), (0,6)\\
 &1&(4,0), (3,1), (2,2), (1,3), (0,4)\\
 &2&(2,0), (1,1), (0,2)\\
 &3&(0,0)\\
\br
\end{tabular}
\end{ruledtabular}

%% file: pairalg_tab02.tex
\DeclareRobustCommand{\half}{\ensuremath{\tfrac12}}
\DeclareRobustCommand{\nhalf}{\ensuremath{-\tfrac12}}
\begin{ruledtabular}
\begin{tabular}{rrlll}
\br
$N$ & $v$ & $[\lambda_1,\lambda_2]_{\grpso{5}}$ & \multicolumn{1}{l}{$(v_1,v_2)$} & \multicolumn{1}{l}{$[\lambda_1,\lambda_2]_{\grpso{4}}$}\\
\mr
0&0&[0,0]& (0,0)& [0,0]\\
1&1&[\half,\half]&(1,0), (0,1)&[\half,\half], [\half,\nhalf]\\
2&0&[0,0]& (0,0)& [0,0]\\
 &2&[1,0]& (1,1), (0,0) & [1,0], [0,0]\\
3&1&[\half,\half]&(1,0), (0,1)&[\half,\half], [\half,\nhalf]\\
4&0&[0,0]& (0,0)& [0,0]\\
\br
\end{tabular}
\end{ruledtabular}

%% file: pairalg_tab03.tex
\begin{ruledtabular}
\begin{tabular}{rrl}
\br
$v$ & $n_{12}-v$ & $(v_1,v_2)$ \\
\mr
0&&
(0,0)\\
2&&
(2,0), (1,1), (0,2), (0,0)\\
4&&
(4,0), (3,1), (2,2), (1,3), (0,4), (2,0), (1,1), (0,2), (0,0)\\
6&4&
(5,1), (4,2), (3,3), (2,4), (1,5), (4,0), (3,1), (2,2), (1,3), (0,4),\\
&& (2,0), (1,1), (0,2), (0,0)\\
8&2&
(5,3), (4,4), (3,5), (4,2), (3,3), (2,4), (3,1), (2,2), (1,3), \\
&&(2,0), (1,1), (0,2), (0,0)\\
10&0&
(5,5), (4,4), (3,3), (2,2), (1,1), (0,0)\\
\br
\end{tabular}
\end{ruledtabular}

%% file: pairalg_tab04.tex
\begin{ruledtabular}
\begin{tabular}{lll}
\br
$E^{(e)}$ & $F^{(f)}$ & \multicolumn{1}{c}{$[E^{(e)},F^{(f)}]^{(g)}$} \\
\mr
$\Gop{aa}{e}$ &
$\Gop{aa}{f}$ &
$(-)^g [1-(-)^{e+f+g}] \hat{e} \hat{f}\smallsixj{e}{f}{g}{j_a}{j_a}{j_a}\Gop{aa}{g}$ 
\\
$\Gop{bb}{e}$ &
$\Gop{bb}{f}$ &
$(-)^g [1-(-)^{e+f+g}] \hat{e} \hat{f}\smallsixj{e}{f}{g}{j_b}{j_b}{j_b}\Gop{bb}{g}$
\\
$\Gop{ab}{e}$ &
$\Gop{ab}{f}$ &
0 
\\
$\Gop{ba}{e}$ & 
$\Gop{ba}{f}$ & 
0 
\\
$\Gop{aa}{e}$ &
$\Gop{bb}{f}$ &
0 
\\
$\Gop{ba}{e}$ &
$\Gop{ab}{f}$ &
$-(-)^{e+f}\hat{e}\hat{f}\smallsixj{e}{f}{g}{j_a}{j_a}{j_b}\Gop{aa}{g}
+(-)^{g}\hat{e}\hat{f}\smallsixj{e}{f}{g}{j_b}{j_b}{j_a}\Gop{bb}{g}$
\\
$\Gop{aa}{e}$ &
$\Gop{ab}{f}$ &
$\theta(-)^{j_a+j_b+g}\hat{e}\hat{f}\smallsixj{e}{f}{g}{j_b}{j_a}{j_a}\Gop{ab}{g}$
\\
$\Gop{aa}{e}$ &
$\Gop{ba}{f}$ &
$-\theta(-)^{j_a+j_b+e+f}\hat{e}\hat{f}\smallsixj{e}{f}{g}{j_b}{j_a}{j_a}\Gop{ba}{g}$
\\
$\Gop{bb}{e}$ &
$\Gop{ab}{f}$ &
$-\theta(-)^{j_a+j_b+e+f}\hat{e}\hat{f}\smallsixj{e}{f}{g}{j_a}{j_b}{j_b}\Gop{ab}{g}$
\\
$\Gop{bb}{e}$ &
$\Gop{ba}{f}$ &
$\theta(-)^{j_a+j_b+g}\hat{e}\hat{f}\smallsixj{e}{f}{g}{j_a}{j_b}{j_b}\Gop{ba}{g}$\\
\br
\end{tabular}
\end{ruledtabular}

%% file: pairalg_tab05.tex
\begin{ruledtabular}
\begin{tabular}{lllr}
\br
$E^{(e)}$ & $F^{(f)}$ & \multicolumn{1}{c}{$[E^{(e)},F^{(f)}]^{(g)}$}  &
\\
\mr
$\Gop{aa}{e}$ &
$\Gop{aa}{f}$ &
$2 (-)^g \hat{e} \hat{f}\smallsixj{e}{f}{g}{j_a}{j_a}{j_a}\Gop{aa}{g}$ 
\\
$\Gop{bb}{e}$ & 
$\Gop{bb}{f}$ & 
$2 (-)^g \hat{e} \hat{f}\smallsixj{e}{f}{g}{j_b}{j_b}{j_b}\Gop{bb}{g}$ 
\\
$\Fop{e}$ & 
$\Fop{f}$ & 
$-2(-)^{j_a+j_b+f} \hat{e} \hat{f} \smallsixj{e}{f}{g}{j_a}{j_a}{j_b}\Gop{aa}{g}
-2(-)^{j_a+j_b+e} \hat{e} \hat{f} \smallsixj{e}{f}{g}{j_b}{j_b}{j_a}\Gop{bb}{g}$ 
\\
$\Gop{aa}{e}$ &
$\Gop{bb}{f}$ &
0 
\\
$\Gop{aa}{e}$ &
$\Fop{f}$ &
$\theta(-)^{j_a+j_b+g} \hat{e} \hat{f} \smallsixj{e}{f}{g}{j_b}{j_a}{j_a}\Fop{g}$
\\
$\Gop{bb}{e}$ &
$\Fop{f}$ &
$\theta(-)^{j_a+j_b+f} \hat{e} \hat{f} \smallsixj{e}{f}{g}{j_a}{j_b}{j_b}\Fop{g}$\\
\br
\end{tabular}
\end{ruledtabular}

%% file: pairalg_tab06.tex
\begin{indented}\item[]
\begin{tabular}{llllll}
\br
%%Algebra &  &$\langle C_2 \rangle_{[\lambda_1\lambda_2\ldots]}$ & Irrep & $\langle C_2 \rangle$
\multicolumn{2}{l}{Algebra} &$\langle C_2 \rangle_{[\lambda_1\lambda_2\ldots]}$ & Irrep & $\langle C_2 \rangle$ & System
\\
\mr
$\grpso{n}$ & $n=2k+1$ & $\sum_{i=1}^k 2\lambda_i(\lambda_i+2k+1-2i)$ &
$[v]$ & $2v(v+n-2)$ & Bosonic single-level\\
$\grpso{n}$ & $n=2k$ & $\sum_{i=1}^k 2\lambda_i(\lambda_i+2k-2i)$ &
$[v]$ & $2v(v+n-2)$ & Bosonic two-level\\
$\grpsp{n}$ & $n=2k$ & $\sum_{i=1}^k 2\lambda_i(\lambda_i+2k+2-2i)$ &
$\lbrace v\rbrace$ & $2v(-v+n+2)$ & Fermionic \\
$\grpu{n}$ & & $\sum_{i=1}^n \lambda_i(\lambda_i+n+1-2i)$ &
$[N]$ & $N(N+n-1)$ & Bosonic \\
&&&
$\lbrace N \rbrace$ & $N(-N+n+1)$ & Fermionic\\
\br
\end{tabular}
\end{indented}

%% file: pairalg.bbl
\newcommand{\identity}[1]{{#1}}
\providecommand{\newblock}{}